\documentclass[12pt]{article}
\usepackage{amssymb,amscd,array}
\catcode `\@=11
\@addtoreset{equation}{section}

\newtheorem{thm}{Theorem}[section]
\newtheorem{rem}[thm]{Remark}
\newtheorem{lemma}[thm]{Lemma}
\newtheorem{prop}[thm]{Proposition}
\newtheorem{cor}[thm]{Corollary}

\def\qed{\blacksquare}
\newcommand{\be}{\begin{equation}}
\newcommand{\ee}{\end{equation}}
\newcommand{\bea}{\begin{eqnarray}}
\newcommand{\eea}{\end{eqnarray}}
\newcommand{\R}{\mathbb{R}}
\newcommand{\N}{\mathbb{N}}
\newcommand{\C}{\mathbb{C}}

\begin{document}
\begin{titlepage}
\begin{center}
{\bf \Large{On the Quantization of the Gravitational Field\\}}
\end{center}
\vskip 1.0truecm
\centerline{D. R. Grigore
\footnote{e-mail: grigore@theor1.theory.nipne.ro, grigore@theory.nipne.ro}}
\vskip5mm
\centerline{Dept. of Theor. Phys., Inst. Atomic Phys.}
\centerline{Bucharest-M\u agurele, P. O. Box MG 6, ROM\^ANIA}
\vskip 2cm
\bigskip \nopagebreak
\begin{abstract}
\noindent
We present a new point of view on the quantization of the gravitational field,
namely we use exclusively the quantum framework of the second quantization.
More explicitly, we take as one-particle Hilbert space, 
${\sf H}_{graviton}$
the unitary irreducible representation of the Poincar\'e group corresponding to
a massless particle of helicity $2$ and apply the second quantization procedure
with Einstein-Bose statistics. The resulting Hilbert space
${\cal F}^{+}({\sf H}_{graviton})$
is, by definition, the Hilbert space of the gravitational field. Then we prove
that this Hilbert space is canonically isomorphic to a space of the type
$Ker(Q)/Im(Q)$
where $Q$ is a supercharge defined in an extension of the Hilbert space 
${\cal F}^{+}({\sf H}_{graviton})$
by the inclusion of ghosts: some Fermion ghosts
$u_{\mu}, \tilde{u}_{\mu}$
which are vector fields and a Bosonic ghost
$\Phi$
which is a scalar field. This has to be contrasted to the usual approaches where
only the Fermion ghosts are considered. However, a rigorous proof that this is,
indeed, possible seems to be laking from the literature.
\end{abstract}
\end{titlepage}

\section{Introduction}

One possible way to perform the quantization of the gravitational field is to
liniarize the classical theory of gravitation using the so called Goldberg
variables \cite{Go}, \cite{Gu1} and then to apply some sort of canonical
quantization of the resulting field theory. Because of the gauge invariance of
the theory (which in this case is the invariance under general coordinates
transformations) one obtains a constrained system and one tries to use a
Bleuler-Gupta type formalism, that's it to start with an Hilbert space endowed
with a sesquilinear non-degenerate form and select the physical states as a
subspace of the type
$Q_{A} \Phi = 0, \quad A = 1,\dots,N$.

Among the pioneering works in this approach we mention \cite{Ti}, \cite{Fa}, 
\cite{OP}, \cite{Gu2}, \cite{KO1}, \cite{KO2}. Using the result of this
analysis many computations have been done in the literature (see \cite{CLM},
\cite{GS}, \cite{Do}, \cite{Za}, \cite{Ve}). 

A related idea is to extend the Fock space to an auxiliary Hilbert space 
${\cal H}^{gh}$ 
including some fictious fields, called ghosts, and construct a supercharge
(i.e. an operator $Q$ verifying $Q^{2} = 0$) such that the physical Hilbert
space is
$
{\cal H}_{phys} \equiv {\it Ker}(Q)/{\it Im}(Q)
$
(see for instance \cite{KO2} and references quoted there).  As a result of this
procedure, it is asserted that the {\it graviton}, i.e. the elementary quantum
particle must be a massless spin $2$ particle. The construction of the
supercharge relies heavily on classical field theory arguments, because one
tries to obtain for the quantum gauge transformations expressions of the same
type as the general coordinates transformations appearing in general
relativity. This invariance is then promoted to a BRST invariance which should
be implemented by the supercommutator with the supercharge.

We will present in this paper a careful analysis of this construction
establishing the equivalence of Hilbert spaces from above in a rigorous way. In
our opinion, this sort of analysis seems to be laking form the literature. In
establishing this result we will use a pure quantum point of view, as advocated
in \cite{SW3}. To be completely rigorous, we define the graviton to be the
unitary irreducible representation of the Poincar\'e group corresponding to
zero mass and helicity $2$ \cite{Va}. The usual Wigner representation for this
elementary system is not very good for practical purposes. To have some analogy
with the situation from the classical field theoretical description of gravity
one has to find out an analogue of the construction of the photon by some
factorization procedure, more precisely to implement a construction of the same
type as that of \cite{Va} ch. IV.7 (see also \cite{SW1}, \cite{Ka}, 
\cite{Gr1}).  The explicit construction is not so elementary as in the case of
the photon and deserves a detailed presentation. One will construct the Hilbert
space of the graviton as a factor space 
$
{\sf H}_{graviton} \equiv {\sf H}'/{\sf H}"
$ 
where 
$
{\sf H}" \subset {\sf H}' \subset {\sf H}
$ 
for some auxiliary Hilbert space 
${\sf H}$, 
as in the case of the photon, but the construction of these Hilbert spaces is
rather complicated as it also is the verification that this factor space
carries the desired representation of the Poincar\'e group. This is done in the
next Section.

Next, we apply the standard second quantization procedures \cite{Co},
\cite{Be}, \cite{Ka}, \cite{Va}
to 
${\sf H}_{graviton}$,
that's it we postulate that the many-gravitons system is the Bosonic Fock space
associated to this one-particle Hilbert space:
${\cal F}_{graviton} \equiv {\cal F}^{+}({\sf H}_{graviton})$.
Like in the case of the photon, one can prove that this Hilbert space can also
be written as a factor space
${\cal F}_{graviton} \equiv {\cal H}'/{\cal H}"$
where 
${\cal H}" \subset {\cal H}' \subset {\cal H}$
for some auxiliary Hilbert space
${\cal H}$.
This is done in Section \ref{sq} and \ref{qgf}.

Next, we prove that one can extend the Hilbert space
${\cal H}$
by including some ghosts: namely, a pair of Fermion ghosts
$u_{\mu}, \tilde{u}_{\mu}$
which are vector fields and a Bosonic ghosts
$\Phi$
which is a scalar field such that we construct a supercharge $Q$ in the
extended Hilbert space
${\cal H}^{gh}$
and prove that
${\cal F}_{graviton} \simeq Ker(Q)/Im(Q)$.
This is done in Section \ref{gh}.

Finally, in Section \ref{gau-obs}, we give the expression for the BRST
transformation and analyse the concept of observables in this framework.

The ``vulnerability" of this type of approach to the construction of free
fields is the fact that, although the ``playground" of the construction, which
is the Fock space, is a canonical object, being canonically constructed from
the one-particle Hilbert space, the expressions of the free fields are deeply
dependent on the representation adopted for this one-particle space. Distinct
factorization procedures (if they do exist) might lead to distinct theories.
It is suggested in the literature \cite{Schr1}, \cite{Schr2} that a road out of
this problem is to use ideas from algebraic field theory. 

We stress here that in the usual approaches only the Fermion ghosts are
considered. However, a rigorous proof that the equality 
${\cal F}_{graviton} \simeq Ker(Q)/Im(Q)$
is true seems to be laking from the literature. We will present at the end of
Section \ref{qgf} some arguments which raises doubts about the correctness of
the standard procedures.

We also mention that a rigorous construction of the Hilbert space of the
many-gravitons system is indispensable for the construction of the
corresponding  $S$-matrix in the sense of perturbation theory of Bogoliubov.
This construction emphasize the basic r\^ole of causality in quantum field
theory (see also \cite{Schr1}, \cite{Schr2} where it is remarked that causality
is again the major physical axiom in algebraic field theory). The recursive
construction of Epstein and Glaser \cite{EG1}, \cite{Gl} as presented in
\cite{Scho1}, \cite{Gri} and {\cite{SW3} makes sense only if this starting
point is settled, because otherwise it is not clear if the elementary physical
particle of the theory is characterized by zero mass and helicity $2$ as should
be the case for the graviton. We will reconsider the construction of a
consistent theory of quantum gravity, in the sense of perturbation theory,
using the formalism developed here in another publication.
\newpage

\section{The Graviton as an Elementary Relativistic \\ Free Particles\label{grav}}

As we have anticipated in the Introduction, one defines the graviton as 
a certain unitary irreducible representation of the Poincar\'e group
corresponding to zero mass and helicity $2$. We will describe this
representation using the formalism of Hilbert space bundles, as presented in
\cite{Va}, ch. VI.7 thm 6.20.

First, we fix some notations. The upper hyperboloid of mass
$m \geq 0$
is by definition
$X^{+}_{m} \equiv \{p \in \R^{4} \vert \quad \Vert p\Vert^{2} = m^{2}\}$;
it is a Borel set with to the Lorentz invariant measure
$d\alpha_{m}^{+}(p) \equiv {d{\bf p} \over 2\omega({\bf p})}$.
Here the conventions are the following:
$\Vert \cdot \Vert$
is the Minkowski norm defined by
$
\Vert p \Vert^{2} \equiv p\cdot p
$
and
$p\cdot q$
is the Minkowski bilinear form:
$
p\cdot q \equiv p_{0}q_{0} - {\bf p} \cdot {\bf q}.
$
If
${\bf p} \in \R^{3}$
we define
$\tau({\bf p}) \in X_{m}^{+}$
according to
$
\tau({\bf p}) \equiv (\omega({\bf p}),{\bf p}), \quad
\omega({\bf p}) \equiv \sqrt{{\bf p}^{2} + m^{2}}.
$

We start now to define the graviton, as a representation of zero mass and
helicity $2$ of the Poincar\'e group. 

\begin{prop}
Let us define by 
${\sf F}$
the space of complex traceless symmetric 
$4 \times 4$
matrices:
\be
{\sf F} \equiv \{ h_{\rho\sigma} \vert h_{\rho\sigma} = h_{\sigma\rho}, 
\quad {h_{\rho}}^{\rho} = 0 \}.
\label{F}
\ee

Then, we define the following objects:

(a) The Borel set:
\be
B \equiv \{ (p_{\mu}, h_{\rho\sigma}) \vert p \in X_{0}^{+}, \quad
h_{\rho\sigma} \in {\sf F}, \quad {h_{\rho}}^{\sigma} p_{\sigma} = 0 \}.
\label{B}
\ee

(b) The canonical projection on the first entry:
$
\pi: B \rightarrow X_{0}^{+};
$

(c) The action of the group
$SL(2,\C)$
on $B$:
\be
D(A) \cdot (p_{\mu}, h_{\rho\sigma} ) \equiv 
({\delta(A)_{\mu}}^{\nu} p_{\nu}, 
{\delta(A)_{\rho}}^{\lambda} {\delta(A)_{\sigma}}^{\omega} h_{\lambda\omega})
\label{D}
\ee
where
$\delta: SL(2,\C) \rightarrow {\cal L}^{\uparrow}_{+}$
is the standard group homomorphism;

(d) The sesquilinear forms
$$
((p,h), (p,h'))_{p} \equiv \overline{h^{\mu\nu}} h'_{\mu\nu}.
$$
(We use the convention of summation over the dummy indices.)

Then
$(X_{0}^{+},B,SL(2,\C),\pi)$
is a pre-Hilbert space bundle. 
\label{pre}
\end{prop}

{\bf Proof:}
One can easily verify that this assertion is correct: 

(i) The operators
$D(A)$
are well define and they give an Borel action of the group
$SL(2,\C)$
on $B$.

(ii) The fibre
$B(p) \equiv \pi^{-1}(\{p\})$
is a vector subspace of $F$ (of dimension $5$).

(iii) The operators
$D(A)$
map
$B(p)$ 
into
$B(\delta(A)\cdot p)$
and leave invariant the sesquilinear forms in the sense:
$$
(D(A)\cdot (p,h), D(A)\cdot (p,h'))_{p} = ((p,h), (p,h'))_{\delta(A)\cdot p}.
$$

(iv) We still have to prove that the sesquilinear forms are positively defined.
For this we use (iii) and the fact that there exists an element 
$A \in SL(2,\C)$
such that
$P \equiv \delta(A)\cdot p$
is of the form
$P^{\mu} = (1,0,0,1)$. 
Let the transformed matrix be
$$
H_{\lambda\omega} \equiv
{\delta(A)_{\rho}}^{\lambda} {\delta(A)_{\sigma}}^{\omega} h_{\lambda\omega}.
$$
One has to write explicitly all the constraints on the matrix $H$, namely the
symmetry, the tracelessness and the property the gives zero when applied to the
quadri-vector $P$. After some elementary computations one discovers that the
generic form of the matrix $H$ is:
\bea
H_{00} = H_{33} = 2F_{0}, \quad H_{03} = H_{30} = - 2F_{0}, 
\nonumber \\
H_{10} = H_{01} = F_{1}, \quad H_{20} = H_{02} = F_{2}, 
\nonumber \\
H_{13} = H_{31} = - F_{1}, \quad H_{23} = H_{32} = - F_{2}, 
\nonumber \\
H_{11} = - H_{22} = \alpha, \quad H_{12} = H_{21} = \beta;
\label{H}
\eea
here the (complex) numbers
$F_{\mu}, \quad \mu = 0,\dots,3$
and
$\alpha, \quad \beta$
are arbitrary. We have now quite elementary
$
\overline{H^{\mu\nu}} H_{\mu\nu} = 2(\alpha^{2} + \beta^{2}) \geq 0
$
which proves the positive definiteness. 
$\qed$

We can easily see that the sesquilinear forms are not strictly
positively defined. We determine in the next proposition the
elements of ``zero norm".
\begin{prop}
Let
$(p,h) \in B$;
then
$((p,h), (p,h))_{p} = 0$
if and only if the matrix $h$ is of the form
$$
h_{\rho\sigma}(p) = p_{\rho} f_{\sigma}(p) + p_{\sigma} f_{\rho}(p)
$$
where the quadri-vector function
$f_{\rho}$
is constrained by the transversality condition:
\be
p^{\rho} f_{\rho}(p) = 0.
\label{trans-f}
\ee
\label{zero}
\end{prop}

{\bf Proof:}
We use the expression $P$ and $H$ defined in the preceding proposition. If we
have 
$((p,h), (p,h))_{p} = 0$
i.e.
$\overline{h^{\mu\nu}} h_{\mu\nu} = 0$
then we also have
$\overline{H^{\mu\nu}} H_{\mu\nu} = 0$.
If we use the relations (\ref{H}) then we obtain
$\alpha = \beta = 0$
and it easily follows that one can write the relations (\ref{H}) in the compact
form
$$
H_{\rho\sigma} = P_{\rho} F_{\sigma} + P_{\sigma} F_{\rho}.
$$
Moreover, we also have
$$
P^{\rho} F_{\rho} = 0.
$$
If we define
$$
f_{\rho} \equiv {\delta(A^{-1})_{\rho}}^{\sigma} F_{\sigma}
$$
then we get the two relations from the statement.
$\qed$

Now we are ready to use the formalism of Hilbert space bundles. 
\begin{prop}
Let us define
\be
{\sf F}_{0} \equiv \{ (p,h) \in {\sf F} 
\vert \overline{h^{\mu\nu}} h_{\mu\nu} = 0 \}, \quad
[{\sf F}] \equiv {\sf F} /{\sf F}_{0}.
\label{FF}
\ee

We construct the following object:

(a) The Borel set:
\be
[B] \equiv \{ (p, [h]) \vert p \in X_{0}^{+}, \quad
[h] \in [{\sf F}], \quad {h_{\rho}}^{\sigma} p_{\sigma} =
0, \quad \forall h \in [h] \}.
\label{BB}
\ee

(b) The canonical projection on the first entry:
$
[\pi]: [B] \rightarrow X_{0}^{+};
$

(c) The action of the group
$SL(2,\C)$
on 
$[B]$:
\be
[D(A)] \cdot (p, [h]) \equiv (\delta(A) \cdot p,
[\delta(A)^{\otimes 2} \cdot h])
\label{DD}
\ee
where we use for simplicity tensor notations (without indices).

(d) The sesquilinear forms
$$
((p,[h]), (p,[h']))_{p} \equiv \overline{h^{\mu\nu}} h'_{\mu\nu}, \quad
\forall h \in [h], \quad \forall h' \in [h'].
$$

Then
$(X_{0}^{+},[B],SL(2,\C),[\pi])$
is a Hilbert space bundle with fibres of dimension $2$. 
\label{Hilbert}
\end{prop}

The proof is elementary: one simply has to check that all objects are well
defined i.e. independent of the choice of the representatives in the
equivalence classes 
$[\cdot]$.

Let us denote the fibre over $p$ by
$[B](p)$.
Then we apply the standard construction from \cite{Va} VI.8 of associating to 
the Hilbert space bundle $[B]$ a representation of the group 
$inSL(2,\C)$
(which is the universal covering group of the proper orthochronous Poincar\'e
group). 
\begin{thm}
Let us construct the vector space
$$
{\cal V} = \{ s: X^{+}_{0} \rightarrow [F] \vert s \quad 
{\rm is~ a~ Borel~ function}, 
\quad s(p) \in [B](p) \}
$$
(i.e. the space of Borel sections of the Hilbert space bundle 
$[B]$)
and define
$$
\Vert s \Vert^{2} \equiv \int_{X^{+}_{0}} d\alpha_{0}^{+}(p) \quad
\left( (p,s(p)),(p, s(p))\right)_{p}.
$$
We define now the space
$$
{\cal V}' \equiv \{ s \in {\cal V} \vert \quad \Vert s \Vert^{2} < \infty \};
$$
then
${\cal V}'$
is a pre-Hilbert space with respect to the scalar product
$$
(s,s') \equiv \int_{X^{+}_{0}} d\alpha_{0}^{+}(p) \quad
\left( (p,s(p)),(p, s'(p))\right)_{p}.
$$
Let 
$\tilde{\cal V}'$
be the Hilbert space which is the completion of 
${\cal V}'$
with respect to the scalar product 
$(\cdot,\cdot)$
and let us define the operators
$U_{a,A}:\tilde{\cal V}' \rightarrow \tilde{\cal V}'$
by
\be
\left( U_{a,A} s\right)(p) \equiv
e^{ia\cdot p} [\delta(A) \cdot h(\delta (A^{-1})\cdot p)],
\quad {\rm for} \quad (a,A) \in inSL(2,\C)
\ee
where 
$h(p) \in s(p)$.
Then $U$ is a unitary representation of 
$inSL(2,\C)$
and it is equivalent to 
the representation
$U^{+,4} \oplus U^{+,-4}$
(see \cite{Va} thm 9.4).
\label{graviton}
\end{thm}

{\bf Proof:}
We want to apply the theorem 6.20 from \cite{Va}. So, we must determine the
action of the stability subgroup
$E^{*}$
on the fibre
$[B](1,0,0,1)$.
We remind that the group
$E^{*}$
is formed from elements of
$SL(2,\C)$
of the following type:
\be
A = \left(\matrix{ z & z^{-1} a \cr 0 & z^{-1} \\}\right), \quad 
\forall z, a \in \C, \quad \vert z \vert = 1.
\ee

We also note that the fibre
$[B](1,0,0,1)$
can be identified as the quotient space of the space of all matrices of the form
(\ref{H}) factorized to the space $V$ of the matrices constrained by the 
condition
$\alpha = \beta = 0$.
It follows that we can identify
$[B](1,0,0,1)$
with the space of matrices
$H_{\alpha,\beta}$
of the form
\be
(H_{\alpha,\beta})_{11} = - (H_{\alpha,\beta})_{22} = \alpha, \quad
(H_{\alpha,\beta})_{12} = (H_{\alpha,\beta})_{21} = \beta, \quad
(H_{\alpha,\beta})_{\rho\sigma} = 0,\quad {\rm in~ rest}.
\label{H0}
\ee
So, we have the isomorphism
$$
[B](1,0,0,1) \ni [H_{\alpha,\beta}] \leftrightarrow 
(\alpha,\beta) \in \C^{2}. 
$$

We have to compute the action of the group 
$E^{*}$
on such elements so we have to compute the matrix
$$
(H'_{\alpha,\beta})_{\rho\sigma} = 
{\delta(A)_{\rho}}^{\lambda} {\delta(A)_{\sigma}}^{\omega} 
(H_{\alpha\beta})_{\lambda\omega}.
$$

This matrix should be of the form
$H_{\alpha'\beta'}(mod(V))$
and one finds out that we have
\bea
\alpha' = \alpha \{ [{\delta(A)_{1}}^{1}]^{2} - [{\delta(A)_{1}}^{2}]^{2} \}
+ 2 \beta {\delta(A)_{1}}^{1} {\delta(A)_{1}}^{2} = 
\Re (z^{4}) \alpha + \Im (z^{4}) \beta,
\nonumber \\
\beta' = \alpha [ {\delta(A)_{1}}^{1} {\delta(A)_{2}}^{1}  -
{\delta(A)_{1}}^{2} {\delta(A)_{2}}^{2}] + 
\beta [ {\delta(A)_{1}}^{1} {\delta(A)_{2}}^{2} +
{\delta(A)_{1}}^{2} {\delta(A)_{2}}^{1} ] =
\nonumber \\
- \Im (z^{4}) \alpha + \Re (z^{4}) \beta.
\eea

In the new variables
$$
u \equiv \alpha + i \beta, \quad v = \alpha - i\beta
$$
we have the transformation rules
$$
u' = z^{-4} u, \quad v' = z^{4} v
$$
i.e. we have the representation
$
\pi_{4} \oplus \pi_{-4}
$
of the stability subgroup
$E^{*}$.
The assertion from the statement follows now from thm. 6.20 of \cite{Va}.
$\qed$

We note that the couple
$(\tilde{\cal V}', U)$
is a unitary representation of the group
$inSL(2,\C)$
which corresponds to zero mass and helicity $2$. According to the usual
physical interpretation, we call this system {\it graviton}. We remark that
this (true) representation of the group
$inSL(2,\C)$
induces a true representation of the group
${\cal P}^{\uparrow}$
as it this the case with all representations of integer spin (or helicity).
Moreover, this representation can be extended to the whole Poincar\'e group.
It is possible to express this representation in an analogous way to the
photon if one considers a factorisation procedure \cite{Va}. Let us consider 
the Hilbert space
${\sf H} \equiv L^{2}(X^{+}_{0},{\sf F},d\alpha_{0}^{+})$
with the scalar product
\be
<\phi,\psi> \equiv \int_{X^{+}_{0}} d\alpha_{0}^{+}(p) \quad
<\phi(p),\psi(p)>_{\sf F}
\ee
where
$$
<\phi,\psi>_{\sf F} \equiv \sum_{\mu,\nu = 0}^{3} 
\overline{\phi_{\mu\nu}} \psi_{\mu\nu}
$$
is the usual scalar product in 
${\sf F}$. 

In this Hilbert space we have the following (non-unitary) representation of 
the Poincar\'e group; we denote by
$I_{s}, \quad I_{t} \in {\cal L}$
the spatial (resp. temporal) inversions.
\bea
\left( U_{a,\Lambda} \phi\right)_{\mu\nu}(p) \equiv
e^{ia\cdot p} {\Lambda_{\mu}}^{\rho}{\Lambda_{\nu}}^{\sigma}
\phi_{\rho\sigma}(\Lambda^{-1}\cdot p)\quad {\rm for} \quad
\Lambda \in {\cal L}^{\uparrow}, 
\nonumber \\
\left( U_{I_{t}} \phi\right)_{\mu\nu}(p) \equiv 
{{I_{t}}_{\mu}}^{\rho}{{I_{t}}_{\nu}}^{\sigma}
\overline{ \phi_{\rho\sigma}(I_{s}\cdot p)}.
\label{reps0}
\eea

Let us define on ${\sf H}$ the following non-degenerate sesquilinear form:
\be
(\phi,\psi) \equiv \int_{X^{+}_{0}} d\alpha_{0}^{+}(p) \quad
\overline{\phi^{\mu\nu}}(p) \psi_{\mu\nu}(p) = 
\int_{X^{+}_{0}} d\alpha_{0}^{+}(p) \quad g^{\mu\nu}
\overline{\phi_{\mu}(p)}\psi_{\nu}(p) = \quad
<\phi,g^{\otimes 2} \psi>;
\ee
here
$g \in {\cal L}^{\uparrow}$
is the Minkowski matrix with diagonal elements
$1,-1,-1,-1$.

Then one easily establishes that we have
\be
(U_{a,\Lambda} \phi,U_{a,\Lambda} \psi) = (\phi,\psi), \quad {\rm for} \quad
\Lambda \in {\cal L}^{\uparrow}, \qquad
(U_{I_{t}} \phi,U_{I_{t}} \psi) = \overline{(\phi,\psi)}..
\label{reprezentation}
\ee

We have now two elementary results:
\begin{lemma}
Let us consider the following subspace of ${\sf H}$:
\be
{\sf H}'\equiv \{ \phi \in {\sf H} \vert \quad p^{\mu} \phi_{\mu\nu}(p) = 0\}.
\ee

Then the sesquilinear form
$\left. (\cdot,\cdot)\right|_{{\sf H}'}$
is positively defined.
\label{h'}
\end{lemma}

We denote 
$\Vert \phi \Vert^{2} \equiv (\phi,\phi).$

\begin{lemma}
Let us consider the following subspace of ${\sf H}'$:
\be
{\sf H}''\equiv \{ \phi \in {\sf H}' \vert \quad \Vert\phi\Vert = 0\}.
\ee

Then the elements of
$
{\sf H}''
$
are of the form
\be
\phi_{\mu\nu}(p) = p_{\mu} f_{\nu}(p) + p_{\nu} f_{\mu}(p)
\ee
where 
$
f:X^{+}_{0} \rightarrow \C^{4}
$
is a Borel function verifying the transversality condition:
$$
p^{\mu} f_{\mu}(p) = 0.
$$
\label{h''}
\end{lemma}

Then we have the following result:
\begin{prop}
The representation (\ref{reprezentation}) of the Poincar\'e group leaves
invariant the subspaces ${\sf H}'$ and ${\sf H}''$ and so, it induces an
representation in the Hilbert space
\be
{\rm H}_{graviton} \equiv \overline{\left( {\sf H}' / {\sf H}''\right)}
\label{graviton1}
\ee
(here by the overline we understand completion). The factor representation,
denoted also by $U$ is unitary and irreducible. By restriction to the proper
orthochronous Poincar\'e group it is equivalent to the representation from the 
preceding theorem.
\end{prop}

\newpage
\section{Second Quantization\label{sq}}

Here we give the main concepts and formul\ae~ connected to the
method of second quantization. We follow essentially \cite{Va}
ch. VII (see also \cite{Co} and \cite{Be}).

The idea of the method of second quantization is to provide a canonical
framework for a multi-particle system in case one has a Hilbert space
describing an ``elementary" particle. One usually takes the one-particle
Hilbert space
${\sf H}$
to be some projective unitary irreducible representation of the Poincar\'e
group. Let
${\sf H}$
be a (complex) Hilbert space; the scalar product on {\sf H} is denoted by
$<\cdot,\cdot>$.
One first considers the tensor algebra
\be
T({\sf H}) \equiv \oplus_{n=0}^{\infty} {\sf H}^{\otimes n},
\ee
where, by definition, the term corresponding to
$n = 0$
is the division field $\C$.
The generic element of
$T({\sf H})$
is of the type
$
(c,\Phi^{(1)},\cdots,\Phi^{(n)},\cdots), \Phi^{(n)} \in {\sf H}^{\otimes n};
$
the element
$\Phi_{0} \equiv (1,0,\dots)$
is called {\it the vacuum}.
Let us consider now the symmetrisation (resp. antisymmetrisation) operators
${\cal S}^{\pm}$
defined by
\be
{\cal S}^{\pm} \equiv \oplus_{n=0}^{\infty} {\cal S}_{n}^{\pm}
\ee
where
${\cal S}_{0}^{\pm} = 1$
and
${\cal S}_{n}^{\pm}, \quad n \geq 1$
are defined on decomposable elements in the usual way
\be
{\cal S}_{n}^{\pm} \phi_{1} \otimes \cdots \otimes \phi_{n} \equiv
{1\over n!} \sum_{P \in {\cal P}_{n}}
\epsilon_{\pm}(P) \phi_{P(1)} \otimes \cdots \otimes \phi_{P(n)}.
\ee
Here
${\cal P}_{n}$
is the group of permutation of the numbers
$1,2,\dots,n$,
$|P|$
is the sign of the permutation $P$ and
$
\epsilon_{+}(P) = 1, \quad \epsilon_{-}(P) = (-1)^{\vert P\vert}
$ 
are the one-dimensional representations of 
${\cal P}_{n}$.

One extends the operators
${\cal S}_{n}^{\pm}$
to arbitrary elements of $T$ by linearity and continuity; it is convenient to
denote the elements in defined by these relations by
$\phi_{1} \vee \cdots \vee \phi_{n}$
and respectively by
$\phi_{1} \wedge \cdots \wedge \phi_{n}$.

We now define the
{\it Bosonic} (resp. {\it Fermionic}) {\it Fock space} according to:
\be
{\cal F}^{\pm}({\sf H}) \equiv {\cal S}^{\pm} T({\sf H});
\ee
obviously we have:
\be
{\cal F}^{\pm}({\sf H}) = \oplus_{n=0}^{\infty} {\cal H}^{\pm}_{n}
\ee
where
\be
{\cal H}^{\pm}_{0} \equiv \C, \quad
{\cal H}^{\pm}_{n} \equiv {\cal S}_{n}^{\pm} {\sf H}^{\otimes n}
\quad (n \geq 1)
\ee
are the so-called $n^{\rm th}$-{\it particle subspaces}.

The operations $\vee$ (resp. $\wedge$) make
${\cal F}^{\pm}({\sf H})$
into associative algebras. One defines in the Bosonic (resp. Fermionic) Fock
space the {\it creation} and {\it annihilation} operators as follow: let
$\phi \in {\sf H}$ be arbitrary.
In the Bosonic case they are defined on elements from
$\psi \in {\cal H}^{+}_{n}$ by
\be
A(\phi)^{\dagger} \psi \equiv \sqrt{n+1} \phi \vee \psi
\ee
and respectively
\be
A(\phi) \psi \equiv {1 \over \sqrt{n}} i_{\phi} \psi
\ee
where
$i_{\phi}$
is the unique derivation of the algebra
${\cal F}^{+}({\sf H})$
verifying
\be
i_{\phi} 1 = 0; \quad i_{\phi} \psi = <\phi,\psi> {\bf 1}.
\ee

\begin{rem}
We note that the general idea is to associate to every element of the
one-particle space
$\phi \in {\sf H}$
a couple of operators
$A^{\sharp}(\phi)$
acting in the Fock space
${\cal F}^{+}({\sf H})$.
\label{creation+annihilation}
\end{rem}

As usual, we have the canonical commutation relations (CCR):
\be
\left[ A(\phi), A(\psi)\right] = 0,\quad
\left[ A(\phi)^{\dagger}, A(\psi)^{\dagger}\right] = 0, \quad
\left[ A(\phi), A(\psi)^{\dagger}\right] = <\phi,\psi> {\bf 1}.
\label{CCR}
\ee

The operators
$A(\psi), \quad A(\psi)^{\dagger}$
are unbounded and adjoint one to the other.

In the Fermionic case we define these operators on elements from
$\psi \in {\cal H}^{-}_{n}$ by
\be
A(\phi)^{\dagger} \psi \equiv \sqrt{n+1} \phi \wedge \psi
\ee
and respectively
\be
A(\phi) \psi \equiv {1 \over \sqrt{n}} i_{\phi} \psi
\ee
where
$i_{\phi}$
is the unique graded derivation of the algebra
${\cal F}^{-}({\sf H})$
verifying
\be
i_{\phi} 1 = 0; \quad i_{\phi} \psi = <\phi,\psi> {\bf 1}.
\ee

Now we have the canonical anticommutation relations (CAR):
\be
\left\{ A(\phi), A(\psi)\right\} = 0,\quad
\left\{ A(\phi)^{\dagger}, A(\psi)^{\dagger}\right\} = 0, \quad
\left\{ A(\phi), A(\psi)^{\dagger}\right\} = <\phi,\psi> {\bf 1}.
\label{CAR}
\ee

The operators
$A(\psi), \quad A(\psi)^{\dagger}$
are bounded and adjoint one to the other.

If $U$ is a unitary (or antiunitary) operator on ${\sf H}$,
it lifts naturally to an operator
$\Gamma(U)$
on the tensor algebra
$T({\sf H})$,
according to
\be
\Gamma(U) \equiv \oplus_{0}^{\infty} U^{\otimes n}
\ee
or, more explicitly, on decomposable elements
\be
\Gamma(U) (\psi_{1} \otimes \cdots \otimes \psi_{n}) =
U \psi_{1} \otimes \cdots \otimes U \psi_{n}.
\ee

The operator
$\Gamma(U)$
leaves invariant the symmetric and resp. the antisymmetric algebras
${\cal F}^{\pm}({\sf H})$
and we have
\be
\Gamma(U) A(\phi) \Gamma(U^{-1}) = A(U\phi).
\label{repsA}
\ee

If $A$ is an self-adjoint operator on 
${\sf H}$
then we define on
$T({\sf H})$
the self-adjoint operator 
$d\Gamma(A)$
according to
\be
d\Gamma(A) (\psi_{1} \otimes \cdots \otimes \psi_{n}) =
A \psi_{1} \otimes \cdots \otimes \psi_{n} + \cdots 
\psi_{1} \otimes \cdots \otimes A \psi_{n},
\ee
continuity and linearity; 
$d\Gamma(A)$
leaves 
${\cal F}^{\pm}({\sf H})$
invariant.
\newpage

\section{The Quantization of the Gravitational Field\label{qgf}}

In this Section we  apply the prescription from Section \ref{sq} to the 
Hilbert space of the graviton
${\rm H}_{graviton}$
given by (\ref{graviton}). The idea is similar to the case of the photon,
namely to express the (Bosonic) {\it Fock space of the graviton}
\be
{\cal F}_{graviton} \equiv {\cal F}^{+}({\rm H}_{graviton})
\ee
as a factorization of the type (\ref{graviton}). It is natural to start with 
the ``bigger"  Fock space
\be
{\cal H} \equiv {\cal F}^{+}({\rm H}) \equiv \oplus_{n\geq 0} {\cal H}_{n}
\label{aux-grav}
\ee
where the
$n^{th}$-particle subspace
${\cal H}_{n}$
is the set of Borel functions
$\Phi^{(n)}_{\mu_{1},\nu_{1};\dots;\mu_{n},\nu_{n}}: (X^{+}_{0})^{\times n}
\rightarrow \C$
which are square summable:
\be
\int_{(X^{+}_{0})^{\times n}} \prod_{i=1}^{n} d\alpha^{+}_{0}(k_{i})
\sum_{\mu_{1},\dots,\mu_{n} =0}^{3}\sum_{\nu_{1},\dots,\nu_{n} =0}^{3}
|\Phi^{(n)}_{\mu_{1},\nu_{1};\dots;\mu_{n},\nu_{n}}(k_{1},\dots,k_{n})|^{2} 
< \infty
\ee
verify the following properties:

(a) symmetry in the triplets
$(k_{i},\mu_{i},\nu_{i}), \quad i = 1,\dots,n$;

(b) symmetry in every couple 
$(\mu_{i},\nu_{i}), \quad i = 1,\dots,n$;

(c) tracelessness in every couple
$(\mu_{i},\nu_{i}), \quad i = 1,\dots,n$.

In ${\cal H}$ the expression of the scalar product is:
\bea
<\Psi,\Phi> \equiv \overline{\Psi^{(0)}} \Phi^{(0)} + \sum_{n=1}^{\infty}
\int_{(X^{+}_{0})^{\times n}} \prod_{i=1}^{n} d\alpha^{+}_{0}(k_{i})
\sum_{\mu_{1},\dots,\mu_{n} =0}^{3}\sum_{\nu_{1},\dots,\nu_{n} =0}^{3}
\nonumber \\
\overline{\Psi^{(n)}_{\mu_{1},\nu_{1};\dots;\mu_{n},\nu_{n}}
(k_{1},\dots,k_{n})}
\Phi^{(n)}_{\mu_{1},\nu_{1};\dots;\mu_{n},\nu_{n}}(k_{1},\dots,k_{n})
\label{scalar-prod1}
\eea
and we have a (non-unitary) representation of the Poincar\'e group given by:
\be
{\cal U}_{g} \equiv \Gamma(U_{g}), \quad \forall g \in {\cal P};
\label{reps1}
\ee
here $U_{g}$ is given by (\ref{reps0}).

Let us define the following non-degenerate sesquilinear form on ${\cal H}$:
\bea
(\Psi,\Phi) \equiv \overline{\Psi^{(0)}} \Phi^{(0)} + \sum_{n=1}^{\infty}
(-1)^{n} \int_{(X^{+}_{0})^{\times n}} \prod_{i=1}^{n}
\left[d\alpha^{+}_{0}(k_{i}) g^{\mu_{i}\rho_{i}}g^{\nu_{i}\sigma_{i}} \right]
\nonumber \\
\overline{\Psi^{(n)}_{\mu_{1},\rho_{1};\dots;\mu_{n},\rho_{n}}
(k_{1},\dots,k_{n})}
\Phi^{(n)}_{\nu_{1},\sigma_{1};\dots;\nu_{n},\sigma_{n}}(k_{1},\dots,k_{n}).
\label{sesq1}
\eea

Then the sesquilinear form
$(\cdot,\cdot)$
behaves naturally with respect to the action of the Poincar\'e group:
\be
({\cal U}_{g}\Psi,{\cal U}_{g}\Phi) = (\Psi,\Phi),
\quad \forall g \in {\cal P}^{\uparrow}, \quad
({\cal U}_{I_{t}}\Psi,{\cal U}_{I_{t}}\Phi) = \overline{(\Psi,\Phi)}.
\ee

We denote
$|\Phi|^{2} = <\Phi,\Phi>$
and
$\Vert\Phi\Vert^{2} = (\Phi,\Phi)$.

Now one has from lemma \ref{h'}:
\begin{lemma}
Let us consider the following subspace of ${\cal H}$:
\be
{\cal H}' \equiv {\cal F}^{+}({\rm H}') = \oplus_{n\geq 0} {\cal H}_{n}'.
\ee

Then
${\cal H}_{n}', \quad n \geq 1$
is generated by elements of the form
$
\phi_{1} \vee \cdots \vee \phi_{n}, \quad
\phi_{1},\dots,\phi_{n} \in {\rm H}'
$
and, in the representation adopted previously for the Hilbert space
${\cal H}_{n}$
we can take
\be
{\cal H}_{n}' = \{ \Phi^{(n)} \in {\cal H}_{n} | \quad k_{1}^{\mu_{1}} 
\Phi^{(n)}_{\mu_{1},\nu_{1};\dots;\mu_{n},\nu_{n}}(k_{1},\dots,k_{n}) = 0\}.
\label{h-prim}
\ee

Moreover, the sesquilinear form
$\left. (\cdot,\cdot)\right|_{{\cal H}'}$
is positively defined.
\label{h'1}
\end{lemma}

Next, one has the analogue of lemma \ref{h''}:
\begin{lemma}

Let ${\cal H}'' \subset {\cal H}'$
given by
\be
{\cal H}'' \equiv \{\Phi \in {\cal H}' | \quad \Vert \Phi \Vert^{2} = 0\} =
\oplus_{n\geq 0} {\cal H}''_{n}
\ee

Then, the subspace
${\cal H}''_{n}, \quad n \geq 1$
is generated by elements of the type
$
\phi_{1} \vee \cdots \vee \phi_{n}
$
where at least one of the vectors
$\phi_{1},\dots,\phi_{n} \in {\rm H}'$
belongs to
${\rm H}''$.

Moreover, in the representation adopted previously for the Hilbert space
${\cal H}_{n}$
the elements of
${\cal H}''_{n}$
are linearly generated by functions of the type:
\bea
\Phi^{(n)}_{\mu_{1},\nu_{1};\dots;\mu_{n},\nu_{n}}(k_{1},\dots,k_{n}) =
{1\over n} \sum_{i=1}^{n} 
[(k_{i})_{\mu_{i}} f_{\nu_{i}}(k_{i}) + (k_{i})_{\nu_{i}} f_{\mu_{i}}(k_{i})]
\times \nonumber \\
\Psi^{(n-1)}_{\mu_{1},\nu_{1};\dots,\hat{\mu_{i}},\hat{\nu_{i}};\dots;
\mu_{n},\nu_{n}}(k_{1},\dots,\hat{k_{i}},\dots,k_{n})
\label{h-secund}
\eea
with
$\Psi \in {\cal H}'$
and
$f: X^{+}_{0} \rightarrow \C^{4}$
verifying the transversality condition:
$k^{\mu} f_{\mu}(k) = 0.$
\label{h''2}
\end{lemma}

Finally we have:
\begin{prop}
There exists an canonical isomorphism of Hilbert spaces
\be
{\cal F}_{graviton} \simeq \overline{{\cal H}'/{\cal H}''}.
\ee
\label{graviton-factor}
\end{prop}

{\bf Proof:}
If
$\psi \in {\rm H}'$
then we denote its class with respect to
${\rm H}''$
by
$[\psi]$;
similarly, if
$\Phi \in {\cal H}'$
we denote its class with respect to
${\cal H}''$
by
$[\Phi].$
Then the application
$
i: \overline{{\cal H}'/{\cal H}''} \rightarrow {\cal F}_{graviton}
$
is well defined by linearity, continuity and
\be
i([\phi_{1} \vee \cdots \vee \phi_{n}]) \equiv
[\phi_{1}] \vee \cdots \vee [\phi_{n}]
\ee
and it is the desired isomorphism. Moreover, the sesquilinear form
$(\cdot,\cdot)$
is strictly positive defined on the factor space, so it induces a scalar
product.
$\qed$

Now we can define the gravitational field as an operator on the Hilbert space
${\cal H}$
in analogy to the construction used for the scalar field (see \cite{Gr1}).

We define for every
$p \in X^{+}_{0}$
the annihilation and creation operators according to:
\be
\left( h_{\rho\sigma}(p) 
\Phi\right)^{(n)}_{\mu_{1},\nu_{1};\dots;\mu_{n},\nu_{n}}
(k_{1},\dots,k_{n}) \equiv \sqrt{n+1} \quad
\Phi^{(n+1)}_{\rho,\sigma;\mu_{1},\nu_{1};\dots;\mu_{n},\nu_{n}}
(p,k_{1},\dots,k_{n}), 
\quad \forall n \in \N
\label{annihilation-grav}
\ee
and
\bea
\left( h^{\dagger}_{\rho\sigma}(p) 
\Phi\right)^{(n)}_{\mu_{1},\nu_{1};\dots;\mu_{n},\nu_{n}}
(k_{1},\dots,k_{n}) \equiv - {\omega({\bf p}) \over \sqrt{n}}
\sum_{i=1}^{n} \delta({\bf p}-{\bf k_{i}}) 
\left( g_{\rho\mu_{i}} g_{\sigma\nu_{i}} + g_{\rho\nu_{i}} g_{\sigma\mu_{i}} 
- {1\over 2} g_{\rho\sigma} g_{\mu_{i}\nu_{i}}\right)
\nonumber \\
\Phi^{(n-1)}_{\mu_{1},\nu_{1};\dots;\hat{\mu_{i}},\hat{\nu_{i}};\dots;
\mu_{n},\nu_{n}} 
(k_{1},\dots,\hat{k_{i}},\dots,k_{n}), 
\qquad \forall n \in \N. \quad
\label{creation-grav}
\eea
We use everywhere consistently the Bourbaki convention:
$
\sum_{\emptyset} \equiv 0.
$
Then one has the {\it canonical commutation relations} (CCR)
\bea
\left[ h_{\rho\sigma}(p), h_{\lambda\omega}(p')\right] = 0,\quad
\left[ h^{\dagger}_{\rho\sigma}(p), h^{\dagger}_{\lambda\omega}(p')\right] = 0,
\nonumber \\
\left[ h_{\rho\sigma}(p), h^{\dagger}_{\lambda\omega}(p')\right] =
-  \omega({\bf p}) \left( g_{\rho\lambda} g_{\sigma\omega} + 
g_{\rho\omega} g_{\sigma\lambda} - 
{1\over 2} g_{\rho\sigma} g_{\lambda\omega}\right)
\delta{({\bf p}-{\bf p}')} {\bf 1}
\label{CCR-grav}
\eea
and the relation
\be
(h^{\dagger}_{\rho\sigma}(p)\Psi,\Phi) = (\Psi,h_{\rho\sigma}(p)\Phi), 
\quad \forall \Psi, \Phi \in {\cal H}
\ee
which shows that
$h_{\rho}^{\dagger}(p)$
is the adjoint of
$h_{\rho}(p)$
with respect to the sesquilinear form
$(\cdot,\cdot)$.

We also have a natural behaviour with respect to the action of the Poincar\'e
group (see (\ref{repsA})):
\bea
{\cal U}_{a,\Lambda} h_{\rho\sigma}(p) {\cal U}^{-1}_{a,\Lambda}  =
e^{ia\cdot p} {(\Lambda^{-1})_{\rho}}^{\lambda} 
{(\Lambda^{-1})_{\sigma}}^{\omega}
h_{\lambda\omega}(\Lambda \cdot p),
\quad \forall \Lambda \in {\cal L}^{\uparrow}, 
\nonumber \\
{\cal U}_{I_{t}} h_{\rho\sigma}(p) {\cal U}^{-1}_{I_{t}}  =
{(I_{t})_{\rho}}^{\lambda} {(I_{t})_{\sigma}}^{\omega}
h_{\lambda\omega}(I_{s} \cdot p)
\label{repsA-em}
\eea
and a similar relation for
$h_{\rho\sigma}^{\dagger}(p)$.

Now we define the {\it gravitational field in the point x} according to
\be
h_{\rho\sigma}(x) \equiv h^{(-)}_{\rho\sigma}(x) + h^{(+)}_{\rho\sigma}(x)
\label{grav1}
\ee
where the expressions appearing in the right hand side are the positive
(negative) frequency parts and are defined by:
\be
h^{(-)}_{\rho\sigma}(x) \equiv {1\over (2\pi)^{3/2}} \int_{X^{+}_{0}}
d\alpha^{+}_{0}(p) e^{ip\cdot x} h_{\rho\sigma}(p),\quad
h^{(+)}_{\rho\sigma}(x) \equiv {1\over (2\pi)^{3/2}} \int_{X^{+}_{0}}
d\alpha^{+}_{0}(p) e^{-ip\cdot x} h^{\dagger}_{\rho\sigma}(p).
\label{grav-pm}
\ee

The explicit expressions are
\be
\left(h^{(-)}_{\rho\sigma}(x) 
\Phi\right)^{(n)}_{\mu_{1},\nu_{1};\dots;\mu_{n},\nu_{n}}
(k_{1},\dots,k_{n}) = {\sqrt{n+1} \over (2\pi)^{3/2}} \int_{X^{+}_{0}}
d\alpha^{+}_{0}(p) e^{ip\cdot x} 
\Phi^{(n+1)}_{\rho,\sigma;\mu_{1},\nu_{1};\dots;\mu_{n},\nu_{n}}
(p,k_{1},\dots,k_{n})
\ee
and
\bea
\left(h^{(+)}_{\rho\sigma}(x) 
\Phi\right)^{(n)}_{\mu_{1},\nu_{1};\dots;\mu_{n},\nu_{n}}
(k_{1},\dots,k_{n}) = - {1\over (2\pi)^{3/2}\sqrt{n}} \sum_{i=1}^{n}
e^{i k_{i}\cdot x} \times
\nonumber \\
\left( g_{\rho\mu_{i}} g_{\sigma\nu_{i}} + 
g_{\rho\nu_{i}} g_{\sigma\mu_{i}} - 
{1\over 2} g_{\rho\sigma} g_{\mu_{i}\nu_{i}}\right)
\Phi^{(n-1)}_{\mu_{1},\nu_{1};\dots;\hat{\mu_{i}},\hat{\nu_{i}};\dots;
\mu_{n},\nu_{n}}(k_{1},\dots,\hat{k_{i}},\dots,k_{n}).
\eea

The properties of the gravitational field operator
$h_{\rho\sigma}(x)$
are summarised in the following elementary proposition:
\begin{prop}
The following relations are true:
\be
(h_{\rho\sigma}(x)\Psi,\Phi) = (\Psi,h_{\rho\sigma}(x)\Phi), \quad \forall
\Psi, \Phi \in {\cal H},
\label{AA-dagger}
\ee
\be
\square h^{(\pm)}_{\rho\sigma}(x) = 0, \quad \square h_{\rho\sigma}(x) = 0,
\label{eqA}
\ee
\be
h_{\rho\sigma}(x) = h_{\sigma\rho}(x), \quad 
{h_{\rho}}^{\rho}(x) = 0
\ee
and
\bea
\left[h^{(\mp)}_{\rho\sigma}(x),h^{(\pm)}_{(\lambda\omega)}(y)\right] =
- {1\over 2} \left( g_{\rho\lambda} g_{\sigma\omega} + 
g_{\rho\omega} g_{\sigma\lambda} - 
{1\over 2} g_{\rho\sigma} g_{\lambda\omega}\right)
D^{(\pm)}_{0}(x-y) \times {\bf 1},
\nonumber \\
\left[h^{(\pm)}_{\rho\sigma}(x),h^{(\pm)}_{\lambda\omega}(y)\right] = 0.
\label{CCR-grav-pm}
\eea
As a consequence we also have:
\be
\left[h_{\rho\sigma}(x),h_{\lambda\omega}(y)\right] = 
- {1\over 2} \left( g_{\rho\lambda} g_{\sigma\omega} + 
g_{\rho\omega} g_{\sigma\lambda} - 
{1\over 2} g_{\rho\sigma} g_{\lambda\omega}\right)
D_{0}(x-y) \times {\bf 1};
\label{commutation-grav}
\ee
here
\be
D_{0}(x) = D_{0}^{(+}(x) + D_{0}^{(-)}(x)
\ee
is the Pauli-Jordan distribution and
$D^{(\pm)}_{0}(x)$
are given by:
\be
D_{0}^{(\pm)}(x) \equiv \pm {1\over (2\pi)^{3/2}} \int_{X^{+}_{0}}
d\alpha^{+}_{0}(p) e^{\mp i p\cdot x}.
\label{D0}
\ee
\end{prop}

One can describe in a convenient way the subspaces
${\cal H}'$
and
${\cal H}''$
using the following operators
\be
L(f) \equiv \int_{\R^{4}} dx f^{\rho}(x) 
\partial^{\sigma} h^{(-)}_{\rho\sigma}(x), \quad
L^{\dagger}(f) \equiv \int_{\R^{4}} dx f^{\rho}(x) 
\partial^{\sigma} h^{(+)}_{\rho\sigma}(x)
\ee
where
$f: \R^{4} \rightarrow \R^{4}$
verify the transversality condition:
${\partial f^{\rho} \over \partial x^{\rho}} = 0$.
Indeed, one has the following result:
\begin{prop}
The following relations are true:
\be
{\cal H}' = \{ \Phi \in {\cal H} | \quad L(f) \Phi = 0,
\quad \forall f \} = \cap_{f} Ker(L(f))
\ee
and
\be
{\cal H}'' = \{ L(f)^{\dagger}\Phi | \quad \forall \Phi \in {\cal H}, \quad
\quad \forall f \} = \cup_{f} Im(L(f)^{\dagger}).
\ee

It follows that we have
\be
{\cal F}_{graviton} = \overline{\cap_{f} Ker(L(f))/
\cup_{f} Im(L(f)^{\dagger})}.
\ee
\label{graviton-fock}
\end{prop}

We construct now some observables on the Fock space of the graviton
${\cal F}_{graviton}$.
As in the case of the photon, we will distinguish a class of observables which
are induced by self-adjoint operators on the Hilbert space
${\cal H}$.
Indeed, if $O$ is such an operator and it leaves invariant the subspaces
${\cal H}'$
and
${\cal H}''$
then it factorizes to an operator on
${\cal F}_{graviton}$
according to the formula
\be
[O] [\Phi] \equiv [O\Phi].
\label{O-inv}
\ee

This type of observables are called {\it gauge invariant observables}. Not all 
observables on
${\cal F}_{graviton}$
are of this type. Now we have the following result:
\begin{lemma}
An operator
$O: {\cal H} \rightarrow {\cal H}$
induces a gauge invariant observable if and only if it verifies:
\be
\left. [L(f), O]\right|_{{\cal H}'} = 0, \quad \forall f.
\label{gauge-inv}
\ee
\label{gi}
\end{lemma}

We end with some comments regarding the perturbative construction of the 
$S$-matrix in the sense of Bogoliubov. Perturbation theory, relies considerably
on the axiom of causality, as shown by H. Epstein and V. Glaser \cite{EG1}.
According to Bogoliubov and Shirkov, the $S$-matrix is constructed inductively
order by order as a formal series of operator valued distributions:
\be
S(g)=1+\sum_{n=1}^\infty{i^{n}\over n!}\int_{\R^{4n}} dx_{1}\cdots dx_{n}\,
T_{n}(x_{1},\cdots, x_{n}) g(x_{1})\cdots g(x_{n}),
\label{S}
\ee
where
$g(x)$
is a tempered test function that switches the interaction and
$T_{n}$
are operator-valued distributions acting in the Fock space of some collection
of free fields; in \cite{EG1} (see also \cite{Gl}) one considers in detail
the case of a real free scalar field. These operator-valued distributions,
which are called {\it chronological products} should verify some properties
which can be argued starting from {\it Bogoliubov axioms}.

$\bullet$
First, it is clear that we can consider them {\it completely symmetrical} in
all variables without loosing generality:
\be
T_{n}(x_{P(1)},\cdots x_{P(n)}) = T_{n}(x_{1},\cdots x_{n}), \quad \forall
P \in {\cal P}_{n}.
\label{sym}
\ee

$\bullet$
Next, we must have {\it Poincar\'e invariance}:
\be
{\cal U}_{a,\Lambda} T_{n}(x_{1},\cdots, x_{n}) {\cal U}^{-1}_{a,\Lambda} =
T_{n}(\Lambda\cdot x_{1}+a,\cdots, \Lambda\cdot x_{n}+a), \quad \forall
\Lambda \in {\cal L}^{\uparrow}.
\label{invariance}
\ee
In particular, {\it translation invariance} is essential for implementing
Epstein-Glaser scheme of renormalisation.

$\bullet$
The central axiom seems to be the requirement of {\it causality} which can be
written compactly as follows. Let us firstly introduce some standard notations.
Denote by
$
V^{+} \equiv \{x \in \R^{4} \vert \quad x^{2} > 0, \quad x_{0} > 0\}
$
and
$
V^{-} \equiv \{x \in \R^{4} \vert \quad x^{2} > 0, \quad x_{0} < 0\}
$
the upper (lower) lightcones and by
$\overline{V^{\pm}}$
their closures. If
$X \equiv \{x_{1},\cdots, x_{m}\} \in \R^{4m}$
and
$Y \equiv \{y_{1},\cdots, y_{n}\} \in \R^{4m}$
are such that
$
x_{i} - y_{j} \not\in \overline{V^{-}}, \quad \forall i=1,\dots,m,\quad
j=1,\dots,n
$
we use the notation
$X \geq Y.$
We use the compact notation
$T_{n}(X) \equiv T_{n}(x_{1},\cdots, x_{n})$
and by
$X \cup Y$
we mean the juxtaposition of the elements of $X$ and $Y$. In particular,
the expression
$T_{n+m}(X \cup Y)$
makes sense because of the symmetry property (\ref{sym}). Then the causality
axiom writes as follows:
\be
T_{n+m}(X \cup Y) = T_{m}(X) T_{n}(Y), \quad \forall X \geq Y.
\label{causality}
\ee

$\bullet$
The {\it unitarity} of the $S$-matrix can be most easily expressed (see
\cite{EG1}) if one introduces, the following formal series:
\be
\bar{S}(g)=1+\sum_{n=1}^\infty{(-i)^{n}\over n!}\int_{\R^{4n}}
dx_{1}\cdots dx_{n}\,\bar{T}_{n}(x_{1},\cdots, x_{n}) g(x_{1})\cdots g(x_{n}),
\label{barS}
\ee
where, by definition:
\be
(-1)^{|X|} \bar{T}_{n}(X) \equiv \sum_{r=1}^{n} (-1)^{r}
\sum_{n_{1}+\cdots +n_{r}=n} \sum_{partitions}
T_{n_{1}}(X_{1})\cdots T_{n_{r}}(X_{r});
\label{antichrono}
\ee
here
$X_{1},\cdots,X_{r}$
is a partition of $X$, $|X|$ is the cardinal of the set $X$ and the sum runs
over all partitions. One calls the operator-valued distributions
$\bar{T_{n}}$
{\it anti-chronological products}. It is not very hard to prove that the series
(\ref{barS}) is the inverse of the series (\ref{S}) i.e. we have:
\be
\bar{S}(g) = S(g)^{-1}
\ee
as formal series. Then the unitarity axiom is:
\be
\bar{T}_{n}(X) = T_{n}(X)^{\dagger}, \quad \forall n \in \N, \quad \forall X.
\label{unitarity}
\ee

$\bullet$
The existence of the {\it adiabatic limit} can be formulated as follows. Let us
take in (\ref{S})
$g \rightarrow g_{\epsilon}$
where
$\epsilon \in \R_{+}$
and
\be
g_{\epsilon}(x) \equiv g(\epsilon x).
\ee

Then one requires that the limit
\be
S \equiv \lim_{\epsilon \searrow 0} S(g_{\epsilon})
\label{adiabatic}
\ee
exists, in the weak sense, and is independent of the the test function $g$. In
other words, the operator $S$ should depend only on the {\it coupling constant}
$g \equiv g(0).$
Equivalently, one requires that the limits
\be
T_{n} \equiv \lim_{\epsilon \searrow 0} T_{n}(g_{\epsilon}^{\otimes n}),
\quad n \geq 1
\label{conservation}
\ee
exists, in the weak sense, and are independent of the test function $g$. One
also calls the limit performed above, the {\it infrared limit}.

$\bullet$
Finally, one demands the {\it stability of the vacuum} i.e.
\be
\lim_{\epsilon \searrow 0} < \Phi_{0},  S(g_{\epsilon}) \Phi_{0}> = 1
\quad \Leftrightarrow \quad 
\lim_{\epsilon \searrow 0}
< \Phi_{0},  T_{n}(g^{\otimes n}_{\epsilon}) \Phi_{0}> = 0,
\quad \forall n \in \N^{*}.
\label{stability}
\ee

A {\it renormalisation theory} is the possibility to construct such a
$S$-matrix starting from the first order term:
\be
T_{1}(x) \equiv {\cal L}(x)
\ee
where
${\cal L}$
is a Wick polynomial called {\it interaction Lagrangian} which should verify
the following axioms:
\be
{\cal U}_{a,\Lambda} {\cal L}(x) {\cal U}^{-1}_{a,\Lambda} =
{\cal L}(\Lambda\cdot x+a),
\quad \forall \Lambda \in {\cal L}^{\uparrow},
\label{inv1}
\ee
\be
\left[{\cal L}(x), {\cal L}(y)\right] = 0, \quad \forall x,y \in \R^{4} \quad
s.t. \quad (x-y)^{2} < 0,
\label{causality1}
\ee
\be
{\cal L}(x)^{\dagger} = {\cal L}(x)
\label{unitarity1}
\ee
and
\be
L \equiv \lim_{\epsilon \searrow 0} {\cal L}(g_{\epsilon})
\label{adiabatic1}
\ee
should exists, in the weak sense, and should be independent of the test
function $g$. Moreover, we should have
\be
<\Phi_{0}, L \Phi_{0}> = 0.
\label{stability1}
\ee

To construct such an operator is not exactly an easy matter. In fact, the set
of relations (\ref{inv1}), (\ref{causality1}) and (\ref{unitarity1}) is a
problem of constructive field theory in the particular case when the Hilbert
space is of the Fock type. In the analysis of Epstein and Glaser \cite{EG1},
\cite{Gl} it is argued that the most natural candidates for interaction
Lagrangian 
${\cal L}(x)$ 
are the Wick polynomials. 

In the familiar physicists language, the problem is to prove that {\it no
ultraviolet divergences} and {\it no infrared divergences} appear, i.e. the
operators
$T_{n}$
are finite and well defined and the adiabatic limit exists. The only freedom
left in this case for a renormalisation theory is the non-uniqueness of the
$T_{n}$'s 
due to {\it finite} normalization terms (also called {\it counterterms}) which 
are distributions with the support 
$
\{x_{1} = \cdots = x_{n} = 0\}.
$ 
The whole construction can be based on the operation of {\it distribution
splitting}; there are very effective ways to perform this operation explicitly
\cite{Sc1}. 

In the case of gravitation, one should modify the preceding axioms of
perturbation theory as follows. One constructs the whole theory in the
auxiliary Hilbert space
${\cal H}$
and imposes the axioms as presented above; then one requires that the resulting
$S$-matrix factorizes, in the adiabatic limit, to the ``physical" Hilbert space
${\cal F}_{graviton}$.  
We will give the explicit form of this {\it factorization axiom} at the end
of the next Section for the case of quantization with ghosts.

If the adiabatic limit do not exists (as it is presumably the case for zero
mass systems as the graviton) one has to consider the factorization axiom as an
heuristic relation and should replace it by another postulate (see \cite{Sc1},
\cite{DHKS1}). 

\newpage

\section{Quantisation with Ghost Fields\label{gh}}

In this subsection we give the complete analysis of another realisation of
${\cal F}_{graviton}$
which is essential for the construction of the perturbation theory in the sense
of the preceding Section. We we exhibit a construction which seems to be new in
the literature and give complete proofs, following the lines of \cite{Gr1}.

First, we give consider the Hilbert space
\be
{\cal H}^{gh} = \sum_{n,m,l,s=0}^{\infty} {\cal H}_{nmls}
\label{Hilbert-ghost}
\ee
where
${\cal H}_{nmls}$
consists of Borel functions
$
\Phi^{(nmls)}_{\mu_{1}\nu_{1},\dots,\mu_{n},\nu_{n};
\rho_{1},\dots,\rho_{m};\sigma_{1},\dots,\sigma_{l}}: 
(X^{+}_{0})^{n+m+l+s} \rightarrow \C
$
such that:

(1) they are  square modulus summable with respect to the Lorentz invariant
measure:
\bea
\sum_{n,m,l,s=0}^{\infty} \int_{(X^{+}_{0})^{n+m+l}} d\alpha^{+}_{0}(K)
d\alpha^{+}_{0}(P) d\alpha^{+}_{0}(Q) d\alpha^{+}_{0}(R)
\nonumber \\
\sum_{\mu_{i},\nu_{i},\rho_{1},\sigma_{i}=0}^{3}
|\Phi^{(nmls)}_{\mu_{1}\nu_{1},\dots,\mu_{n},\nu_{n};
\rho_{1},\dots,\rho_{m};\sigma_{1},\dots,\sigma_{l}}(K;P;Q;R)|^{2} \leq \infty
\eea
where we are using the condensed notations:
$
K \equiv (k_{1},\dots,k_{n}),
$
$
P \equiv (p_{1},\dots,p_{m}),
$
$
Q \equiv (q_{1},\dots,q_{l})
$
and
$
R \equiv (r_{1},\dots,r_{s}).
$

(2) they verify the following (anti)symmetry properties:

(a) symmetry at the permutation of the triplets
$(k_{i},\mu_{i},\nu_{i}) \leftrightarrow (k_{j},\mu_{j},\nu_{j}), 
\quad i,j = 1,\dots,n$;

(b) symmetry at the change
$\mu_{i} \leftrightarrow \nu_{i}, \quad i = 1,\dots,n$;

(c) antisymmetry at the permutation of the couples
$(p_{i},\rho_{i}) \leftrightarrow (p_{j},\rho_{j}), \quad i,j = 1,\dots, m$;

(d) antisymmetry at the permutation of the couples
$(q_{i},\sigma_{i}) \leftrightarrow (q_{j},\sigma_{j}), 
\quad i,j = 1,\dots,l$;

(e) antisymmetry in the variables
$r_{i}, \quad i = 1,\dots,s$.

(3) tracelessness in every couple
$(\mu_{i},\nu_{i}), \quad i = 1,\dots,n$.

In this representation we define the annihilation operators according to the
expressions:
\bea
\left( h_{\alpha\beta}(t) \Phi\right)^{(nmls)}_{\mu_{1}\nu_{1},
\dots,\mu_{n},\nu_{n};\rho_{1},\dots,\rho_{m};\sigma_{1},\dots,\sigma_{l}}
(K;P;Q;R) \equiv
\nonumber \\ \sqrt{n+1}
\Phi^{(n+1,mls)}_{\alpha\beta,\mu_{1}\nu_{1},\dots,\mu_{n},\nu_{n};
\rho_{1},\dots,\rho_{m};\sigma_{1},\dots,\sigma_{l}}(t,K;P;Q;R)
\eea
\bea
\left( b_{\alpha}(t) \Phi\right)^{(nmls)}_{\mu_{1}\nu_{1},
\dots,\mu_{n}\nu_{n};\rho_{1},\dots,\rho_{m};\sigma_{1},\dots,\sigma_{l}}
(K;P;Q;R) \equiv
\nonumber \\ \sqrt{m+1}
\Phi^{(n,m+1,l,s)}_{\mu_{1}\nu_{1},\dots,\mu_{n},\nu_{n};
\alpha,\rho_{1},\dots,\rho_{m};\sigma_{1},\dots,\sigma_{l}}(K;t,P;Q;R)
\eea
\bea
\left( c_{\alpha}(t) \Phi\right)^{(nmls)}_{\mu_{1}\nu_{1},
\dots,\mu_{n},\nu_{n};\rho_{1},\dots,\rho_{m};\sigma_{1},\dots,\sigma_{l}}
(K;P;Q;R) \equiv
\nonumber \\  (-1)^{m} \sqrt{l+1}
\Phi^{(nm,l+1,s)}_{\mu_{1}\nu_{1},\dots,\mu_{n},\nu_{n};
\rho_{1},\dots,\rho_{m};\alpha,\sigma_{1},\dots,\sigma_{l}}(K;P;t,Q;R)
\eea
and
\bea
\left( d(t) \Phi\right)^{(nmls)}_{\mu_{1}\nu_{1},\dots,\mu_{n}\nu_{n};
\rho_{1},\dots,\rho_{m};\sigma_{1},\dots,\sigma_{l}}(K;P;Q;R) \equiv
\nonumber \\ \sqrt{s+1}
\Phi^{(n,m,l,s+1)}_{\mu_{1}\nu_{1},\dots,\mu_{n},\nu_{n};
\rho_{1},\dots,\rho_{m};\sigma_{1},\dots,\sigma_{l}}(K;P;Q;t,R).
\eea

The expressions for the creation operators are:
\bea
\left( h^{\dagger}_{\alpha\beta}(t) \Phi\right)^{(nmls)}_{\mu_{1}\nu_{1},
\dots,\mu_{n},\nu_{n};\rho_{1},\dots,\rho_{m};\sigma_{1},\dots,\sigma_{l}}
(k_{1},\dots,k_{n};P;Q;R) \equiv
\nonumber \\ 
-~ \omega({\bf t}) \sum_{i=1}^{n} \delta({\bf t}-{\bf k_{i}}) 
\left( g_{\alpha\mu_{i}} g_{\beta\nu_{i}} + 
g_{\alpha\nu_{i}} g_{\beta\mu_{i}} 
- {1\over 2} g_{\alpha\beta} g_{\mu_{i}\nu_{i}}\right) \times
\nonumber \\
\Phi^{(n-1,mls)}_{\mu_{1}\nu_{1},\dots,\hat{\mu_{i}}\hat{\nu_{i}},\dots,
\mu_{n},\nu_{n};\rho_{1},\dots,\rho_{m};\sigma_{1},\dots,\sigma_{l}}
(k_{1},\dots,\hat{k_{i}},\dots,k_{n};P;Q;R)
\eea
\bea
\left( b^{*}_{\alpha}(t) \Phi\right)^{(nmls)}_{\mu_{1}\nu_{1},
\dots,\mu_{n}\nu_{n};\rho_{1},\dots,\rho_{m};\sigma_{1},\dots,\sigma_{l}}
(K;p_{1},\dots,p_{m};Q;R) \equiv \omega({\bf t}) \times
\nonumber \\ 
\sum_{i=1}^{m} (-1)^{i-1} \delta({\bf t}-{\bf p_{i}}) 
g_{\alpha\rho_{i}}
\Phi^{(n,m-1,l,s)}_{\mu_{1}\nu_{1},\dots,\mu_{n},\nu_{n};
\rho_{1},\dots,\hat{\rho_{i}},\dots,\rho_{m};\sigma_{1},\dots,\sigma_{l}}
(K;p_{1},\dots,\hat{p_{i}},\dots,p_{m};Q;R)
\eea
\bea
\left( c^{*}_{\alpha}(t) \Phi\right)^{(nmls)}_{\mu_{1}\nu_{1},
\dots,\mu_{n},\nu_{n};\rho_{1},\dots,\rho_{m};\sigma_{1},\dots,\sigma_{l}}
(K;P;q_{1},\dots,q_{l};R) \equiv (-1)^{m} \omega({\bf t}) \times
\nonumber \\  
\sum_{i=1}^{l} (-1)^{i-1} \delta({\bf t}-{\bf q_{i}}) 
g_{\alpha\sigma_{i}}
\Phi^{(nm,l-1,s)}_{\mu_{1}\nu_{1},\dots,\mu_{n},\nu_{n};
\rho_{1},\dots,\rho_{m};\sigma_{1},\dots,\hat{\sigma_{i}},\dots,\sigma_{l}}
(K;P;q_{1},\dots,\hat{q_{i}},\dots,q_{l};R)
\eea
and
\bea
\left( d^{*}(t) \Phi\right)^{(nmls)}_{\mu_{1}\nu_{1},\dots,\mu_{n}\nu_{n};
\rho_{1},\dots,\rho_{m};\sigma_{1},\dots,\sigma_{l}}
(K;P;Q;r_{1},\dots,r_{s}) \equiv \omega({\bf t}) \times
\nonumber \\ 
\sum_{i=1}^{s} \delta({\bf t}-{\bf r_{i}}) 
\Phi^{(n,m,l,s-1)}_{\mu_{1}\nu_{1},\dots,\mu_{n},\nu_{n};
\rho_{1},\dots,\rho_{m};\sigma_{1},\dots,\sigma_{l}}
(K;P;Q;r_{1},\dots,\hat{r_{i}},\dots,r_{s}).
\eea

We note that the Hilbert space 
${\cal H}$
given by (\ref{aux-grav}) can be naturally embedded into
${\cal H}^{gh}$
as follows:
\be
{\cal H} \sim \sum_{n=0}^{\infty} {\cal H}_{n000}.
\ee
Moreover, this embedding preserves the expressions of the operators
$h^{\#}_{\alpha\beta}$.
For this reason we call the Fock space
${\cal H}^{gh}$
{\it the ghost extension} of
${\cal H}$.
Remark the choice of the various statistics which seems to be essential for the
whole analysis.

They verify the canonical (anti)commutation relations (\ref{CCR-grav}) and
\be
\{b_{\alpha}(k),b_{\beta}(q)\} = 0, \quad
\{b^{*}_{\alpha}(k),b^{*}_{\beta}(q)\} = 0, \quad
\{b_{\alpha}(k),b^{*}_{\beta}(q)\} = 2 \omega({\bf q}) g_{\alpha\beta}
\delta({\bf k} - {\bf q}) {\bf 1}
\ee
\be
\{c_{\alpha}(k),c_{\beta}(q)\} = 0, \quad
\{c^{*}_{\alpha}(k),c^{*}_{\beta}(q)\} = 0, \quad
\{c_{\alpha}(k),c^{*}_{\beta}(q)\} = 2 \omega({\bf q}) g_{\alpha\beta}
\delta({\bf k} - {\bf q}) {\bf 1}
\ee
\be
[d(k),d(q)] = 0, \quad [d^{*}(k),d^{*}(q)] = 0, \quad
[d(k),d^{*}(q)] = 2 \omega({\bf q}) \delta({\bf k} - {\bf q}) {\bf 1}
\ee
\be
\{b^{\#}_{\alpha}(k),c^{\#}_{\beta}(q)\} = 0, \quad 
[b^{\#}_{\alpha}(k),d^{\#}(q)] = 0, \quad 
[c^{\#}_{\alpha}(k),d^{\#}(q)] = 0
\ee
\be
[h^{\#}_{\alpha\beta}(k),b^{\#}_{\gamma}(q)] = 0, \quad 
[h^{\#}_{\alpha\beta}(k),c^{\#}_{\gamma}(q)] = 0, \quad 
[h^{\#}_{\alpha\beta}(k),d^{\#}(q)] = 0.
\ee

In this Hilbert space a (non-unitary) representation of the Poincar\'e group
acts in an obvious way and the creation and annihilation operators defined
above behave naturally with respect to these Poincar\'e transformations.

Then we can define, beside the gravitational field (see the preceding Section) 
the following Fermionic fields
\be
u(x) \equiv {1\over (2\pi)^{3/2}} \int_{X^{+}_{0}} d\alpha^{+}_{0}(q)
\left[ e^{-i q\cdot x} b(q) + e^{i q\cdot x} c^{*}(q) \right]
\label{u}
\ee
\be
\tilde{u}(x) \equiv {1\over (2\pi)^{3/2}} \int_{X^{+}_{0}} d\alpha^{+}_{0}(q)
\left[ - e^{-i q\cdot x} c(q) + e^{i q\cdot x} b^{*}(q) \right]
\label{u-tilde}
\ee
and the Bosonic field
\be
\Phi(x) \equiv {1\over (2\pi)^{3/2}} \int_{X^{+}_{0}} d\alpha^{+}_{0}(q)
\left[ e^{-i q\cdot x} d(q) + e^{i q\cdot x} d^{*}(q) \right]
\label{phi}
\ee
which are called {\it ghost fields}. They verify the wave equations:
\be
\square u(x) = 0, \quad \square \tilde{u}(x) = 0, \quad \square \Phi(x) = 0
\label{equ}
\ee
and if we identify, as usual the positive (negative) frequency parts we have
the canonical anticommutation relations:
$$
\{ u_{\mu}^{(\epsilon)}(x),u_{\nu}^{(\epsilon')}(y)\} = 0, \quad
\{u_{\mu}(x),u_{\nu}(y)\} = 0, \quad
\{ \tilde{u}_{\mu}^{(\epsilon)}(x),\tilde{u}_{\nu}^{(\epsilon')}(y)\} = 0, 
\quad \{\tilde{u}_{\mu}(x),\tilde{u}_{\nu}(y)\} = 0,
$$
$$
\{ u_{\mu}^{(\epsilon)}(x),\tilde{u}_{\nu}^{(-\epsilon)}(y)\} = 
g_{\mu\nu} D^{(-\epsilon)}_{0}(x-y) {\bf 1}, \quad 
\{u_{\mu}(x),\tilde{u}_{\nu}(y)\} = D_{0}(x-y) {\bf 1}
$$
$$
[\Phi^{(\epsilon)}(x),\Phi^{(\epsilon)}(y)] = 0, \quad
[\Phi^{(\epsilon)}(x),\Phi^{(-\epsilon)}(y)] = D^{(-\epsilon)}_{0}(x-y) 
{\bf 1}, \quad 
[\Phi(x),\Phi(y)] = D_{0}(x-y) {\bf 1}.
$$
$$
[u_{\mu}^{(\epsilon)}(x),\Phi^{(\epsilon')}(y)] = 0, \quad
[\tilde{u}_{\mu}^{(\epsilon)}(x),\Phi^{(\epsilon')}(y)] = 0.
\quad \forall \epsilon, \epsilon' = \pm. 
$$

Now we can introduce an important operator:
\be
Q \equiv \int_{X^{+}_{0}} d\alpha^{+}_{0}(k) k^{\mu}
\left[ h_{\mu\nu}(k) c^{*\nu}(k) + h^{\dagger}_{\mu\nu}(k) b^{\nu}(k) +
{1\over 2} \left( d(k) c^{*\nu}(k) + d^{*}(k) b^{\nu}(k) \right) \right]
\label{supercharge}
\ee
called {\it supercharge}. Its properties are summarised in the following
proposition which can be proved by elementary computations:
\begin{prop}
The following relations are valid:
\be
Q \Phi_{0} = 0;
\label{Q-0}
\ee
\be
\left[ Q, h^{\dagger}_{\mu\nu}(k) \right] = - {1\over 2} 
\left( g_{\rho\mu} g_{\sigma\nu} + g_{\rho\nu} g_{\sigma\mu} 
- {1\over 2} g_{\rho\sigma} g_{\mu\nu}\right) k^{\rho} c^{*\sigma}(k),
\label{com-dagger1}
\ee
\be
\left\{ Q, b_{\mu}^{*}(k) \right\} = k^{\nu} h^{\dagger}_{\mu\nu}(k)
+ {1\over 2} k_{\mu} d^{*}(k), \quad
\left\{ Q, c_{\mu}^{*}(k) \right\} = 0, \quad
\left[ Q, d^{*}(k)\right] = {1\over 2} k^{\mu} c_{\mu}^{*}(k);
\label{com-dagger2}
\ee
\be
\left[ Q, h_{\mu\nu}(k) \right] = {1\over 2} 
\left( g_{\rho\mu} g_{\sigma\nu} + g_{\rho\nu} g_{\sigma\mu} 
- {1\over 2} g_{\rho\sigma} g_{\mu\nu}\right) k^{\rho} b^{\sigma}(k),
\label{com1}
\ee
\be
\left\{ Q, b_{\mu}(k) \right\} = 0, \quad
\left\{ Q, c_{\mu}(k) \right\} = k^{\nu} h_{\mu\nu}(k) + 
{1\over 2} k_{\mu} d(k) \quad
\left[ Q, d(k) \right] = - {1\over 2} k^{\mu} b_{\mu}(k);
\label{com2}
\ee
\be
Q^{2} = 0;
\label{square}
\ee
\be
Im(Q) \subset Ker(Q)
\label{im-ker}
\ee
and
\be
{\cal U}_{g} Q = Q {\cal U}_{g}, \quad \forall g \in {\cal P}.
\label{UQ}
\ee

Moreover, one can express the supercharge in terms of the ghosts fields as
follows:
\be
Q = \int_{\R^{3}} d^{3}x  \left[ \partial^{\mu} h_{\mu\nu}(x)
\stackrel{\leftrightarrow}{\partial_{0}}u^{\nu}(x) + {1\over 4}
\partial^{\mu} u_{\mu}(x)
\stackrel{\leftrightarrow}{\partial_{0}}\Phi(x)\right] 
\ee
\end{prop}

In particular (\ref{square}) justify the terminology of supercharge and
(\ref{im-ker}) suggests that we should compute the quotient. We will
rigorously prove that this quotient coincides with
${\cal F}_{graviton}$.

Let us also give the explicit expression of the supercharge which will be
needed further; starting from the definition (\ref{supercharge}) we immediately
get:
\bea
\left(Q
\Phi\right)^{(nmls)}_{\mu_{1}\nu_{1},\dots,\mu_{n}\nu_{n};
\rho_{1},\dots,\rho_{m};\sigma_{1},\dots,\sigma_{l}}
(k_{1},\dots,k_{n};p_{1},\dots,p_{m};q_{1},\dots,q_{l};r_{1},\dots,r_{s}) =
(-1)^{m} \sqrt{n+1\over l}
\nonumber \\
\sum_{i=1}^{l} (-1)^{i-1} q^{\alpha}_{i}
\Phi^{(n+1,m,l-1,s)}_{\alpha\sigma_{i},\mu_{1},\nu_{1},\dots,\mu_{n}\nu_{n};
\rho_{1},\dots,\rho_{m};\sigma_{1},\dots,\hat{\sigma_{i}},\dots,\sigma_{l}}
(q_{i},k_{1},\dots,k_{n};P;q_{1},\dots,\hat{q_{i}},\dots,q_{l};R)
\nonumber \\
- {1\over 2} \sqrt{m+1\over n} \sum_{i=1}^{n} 
[(k_{i})_{\mu_{i}} \delta^{\alpha}_{\nu_{i}} +
(k_{i})_{\nu_{i}} \delta^{\alpha}_{\mu_{i}} - {1\over 2}
k_{i}^{\alpha} g_{\mu_{i}\nu_{i}} ] \times
\nonumber \\
\Phi^{(n-1,m+1,l,s)}_{\mu_{1}\nu_{1},\dots,\hat{\mu_{i}}\hat{\nu_{i}},\dots,
\mu_{n}\nu_{n};\rho_{1},\dots,\rho_{m};\sigma_{1},\dots,\sigma_{l}}
(k_{1},\dots,\hat{k_{i}},\dots,k_{n};k_{i},p_{1},\dots,p_{m};Q;R)
\nonumber \\
+ {1\over 4} \sqrt{s+1\over l} \sum_{i=1}^{l} (-1)^{i-1} (q_{i})_{\sigma_{i}}
\times 
\nonumber \\
\Phi^{(n,m,l-1,s+1)}_{\mu_{1},\nu_{1},\dots,\mu_{n}\nu_{n};
\rho_{1},\dots,\rho_{m};\sigma_{1},\dots,\hat{\sigma_{i}},\dots,\sigma_{l}}
(K;P;q_{1},\dots,\hat{q_{i}},\dots,q_{l};q_{i},r_{1},\dots,r_{l})
\nonumber \\
+ {1\over 4} \sqrt{m+1\over s} \sum_{i=1}^{n} 
r_{i}^{\alpha} 
\Phi^{(n,m+1,l,s-1)}_{\mu_{1}\nu_{1},\dots,\mu_{n}\nu_{n};\alpha,
\rho_{1},\dots,\rho_{m};\sigma_{1},\dots,\sigma_{l}}
(K;r_{i},p_{1},\dots,p_{m};Q;r_{1},\dots,r_{l}) \qquad
\label{Q-explicit}
\eea
where, of course, we use Bourbaki convention
$\sum_{\emptyset} \equiv 0$.

\begin{rem}
Let us note that the relations (\ref{com-dagger1}), (\ref{com-dagger2})  and 
(\ref{Q-0}) are {\it uniquely} determining the expression of the supercharge.
\end{rem}

Now we introduce on
${\cal H}^{gh}$
the {\it Krein operator} which in compact tensorial notations looks:
\be
\left( J\Phi\right)^{(nmls)}(K;P;Q;R) \equiv
(-1)^{ml+n} g^{\otimes 2n+m+l} \Phi^{(nlms)}(K;Q;P;R).
\label{Krein-gh}
\ee

The properties of this operator are given below and can be proved by elementary
computations:
\begin{prop}
The following relations are verified:
\be
J^{*} = J^{-1} = J
\ee
\be
J b_{\mu}(p) J = c_{\mu}(p), \quad J d(p) J = d(p), \quad 
J h^{*}_{\mu\nu}(p) J = h^{\dagger}_{\mu\nu}(p)
\ee
\be
J Q J = Q^{*}
\ee
and
\be
{\cal U}_{g} J = J {\cal U}_{g}, \quad \forall g \in {\cal P}.
\label{UJ}
\ee
Here $O^{*}$ is the adjoint of the operator $O$ with respect to the scalar
product
$<\cdot,\cdot>$
on
${\cal H}^{gh}$.
\end{prop}

We can define now the sesquilinear form on
${\cal H}^{gh}$ according to
\be
(\Psi,\Phi) \equiv <\Psi,J\Phi>;
\label{sesqui-gh}
\ee
then this form is non-degenerated.  It is convenient to denote the conjugate of
the arbitrary operator $O$ with respect to the sesquilinear form
$(\cdot,\cdot)$
by
$O^{\dagger}$
i.e.
\be
(O^{\dagger} \Psi,\Phi) = (\Psi,O \Phi).
\ee

Then the following formula is available:
\be
O^{\dagger} = J O^{*} J.
\ee

As a consequence, we have
\be
h_{\mu\nu}(x)^{\dagger} = h_{\mu}(x), \quad
u(x)^{\dagger} = u(x), \quad
\tilde{u}(x)^{\dagger} = - \tilde{u}(x), \quad
\Phi^{\dagger}(x) = \Phi(x).
\label{conjugate}
\ee

From (\ref{UJ}) it follows that we have:
\be
({\cal U}_{g}\Psi,{\cal U}_{g}\Phi) = (\Psi,\Phi),
\forall g \in {\cal P}^{\uparrow}, \quad
({\cal U}_{I_{t}}\Psi,{\cal U}_{I_{t}}\Phi) = \overline{(\Psi,\Phi)}.
\label{inv-sesqui}
\ee

Now, we concentrate on the description of the factor space
$Ker(Q)/Im(Q)$.
The following analysis, which is essential for the understanding of graviton
description seems to be missing from the literature. We will construct a
``homotopy" for the supercharge $Q$. We make an {\it anszatz}:
\begin{prop}
Let us define the operator
\be
\tilde{Q} \equiv \int_{X^{+}_{0}} d\alpha^{+}_{0}(q) \left(
f^{\alpha\beta;\nu}(k)
\left[ h_{\alpha\beta}(k) b^{*}_{\nu}(k) + 
h^{\dagger}_{\alpha\beta}(k) c_{\nu}(k)\right] +
a f^{\nu}(k) \left[c_{\nu}(k) d^{*}(k) + b_{\nu}^{*}(k) d(k)\right] \right)
\label{tildeQ}
\ee
with Borel functions
$f^{\nu}$,
$f^{\alpha\beta;\nu}$
(which is symmetric in 
$\alpha$ 
and
$\beta$)
and
$a$ a real number.

Then one can choose these functions in such a way that the following relation
is valid:
\be
Y \equiv \{Q,\tilde{Q}\} = N_{b} + N_{c} + N_{d} + d\Gamma (P) \otimes {\bf 1}
\otimes {\bf 1} \otimes {\bf 1}
\label{Y}
\ee
where 
$N_{b}$,
$N_{c}$
and
$N_{d}$
are particle number operators for the ghosts of type $b$ (resp. $c$ and $d$) 
and $P$ is the projector operator acting in the one-particle graviton space
according to the a formula of the following type:
\be
(P \psi)_{\mu\nu}(k) \equiv {G_{\mu\nu}}^{\rho\sigma}(k) \psi_{\rho\sigma}(k).
\label{P}
\ee

Moreover the following relations are true:
\be
\tilde{Q}^{2} = 0
\ee
and
\be
[Y, Q] = 0, \quad [Y, \tilde{Q}] = 0.
\label{YQ}
\ee
\end{prop}

{\bf Proof:}
(i) First, one computes the generic expression $Y$ using (\ref{supercharge})
and (\ref{tildeQ}). By elementary computations we get
\bea
Y = \int_{X^{+}_{0}} d\alpha^{+}_{0}(k) 
[ F^{\mu\nu;\alpha\beta}(k) h^{\dagger}_{\mu\nu}(k) h_{\alpha\beta}(k) 
- F^{\mu\nu}(k) ( b^{*}_{\nu}(k) b_{\mu}(k) + c_{\mu}^{*}(k) c_{\nu}(k)) 
\nonumber \\
+ H^{\mu\nu}(k) (d^{*}(k) h_{\mu\nu}(k) + h^{\dagger}_{\mu\nu}(k) d(k)) 
+ a k^{\mu} f_{\mu}(k) d^{*}(k) d(k) ]
\label{Y1}
\eea
where
\be
F^{\mu\nu;\alpha\beta} = {1\over 2}
(f^{\alpha\beta;\nu} k^{\mu} + f^{\alpha\beta;\mu} k^{\nu} + 
f^{\mu\nu;\alpha} k^{\beta} + f^{\mu\nu;\beta} k^{\alpha}),
\ee
\be
F_{\mu\nu} = f_{\mu\rho;\nu} k^{\rho} - {1\over 4} {f^{\alpha}}_{\alpha;\nu} 
k_{\mu} - {a\over 2} f_{\nu} k_{\mu}
\ee
and
\be
H_{\mu\nu} = {1\over 2} [f_{\mu\nu;\alpha} k^{\alpha} + 
a (f_{\nu} k_{\mu} + f_{\mu} k_{\nu})].
\ee

(ii) Next, we use the expressions for 
$h^{\#}_{\mu\nu}$
and prove that the first contribution in (\ref{Y1}) can be rewritten as
$
d\Gamma (P) \otimes {\bf 1} \otimes {\bf 1} \otimes {\bf 1}
$
where the operator $P$ has the expression (\ref{P}) with
\be
G_{\mu\nu;\alpha\beta} = - \left( F_{\mu\nu;\alpha\beta} -
{1\over 4} g_{\mu\nu} {F^{\lambda}}_{\lambda;\alpha\beta}\right).
\ee

(iii) Finally, we must make an convenient anszatz for the functions
$f^{\nu}$
and
$f^{\alpha\beta;\nu}$.
We can find out \cite{Gr1} a function
$f: X^{+}_{0} \rightarrow \C^{4}$
such that
\be
k_{\mu} f^{\mu}(k) = - 1, \quad f^{\mu}(k) f_{\mu}(k) = 0.
\label{eps}
\ee

Then we take 
\be
f_{\rho\sigma;\nu} = f_{\rho} g_{\sigma\nu} + f_{\sigma} g_{\rho\nu} +
b f_{\rho} f_{\sigma} k_{\nu} + c (f_{\rho} k_{\sigma} + k_{\rho} f_{\sigma})
f_{\nu}
\label{anszatz-f}
\ee
with $b$ and $c$ real numbers, and impose the following conditions:
\be
G_{\mu\nu;\alpha\beta} G^{\alpha\beta;\rho\sigma} = 
{G_{\mu\nu}}^{\rho\sigma}
\ee
(which ensures that the operator $P$ given by the expression (\ref{P}) is
indeed a projector) and
\be
F_{\mu\nu} = - g_{\mu\nu}, \quad H_{\mu\nu} = 0.
\ee

By elementary but tedious computations one finds out that these conditions are
fulfilled if we take
$a = - 1$,
$b = 1$
and
$c = 0$.
As a result, the formula (\ref{Y}) for $Y$ from the statement follows. The next
two formul\ae~ from the statement are now the result of some elementary
computations. 
$\qed$

We call the operator
$\tilde{Q}$
the {\it homotopy} of $Q$. We have now
\begin{prop}
The operator
$\left. Y\right|_{{\cal H}_{nmls}}$ is invertible iff
$m + l + s > 0$.
\end{prop}

{\bf Proof:}
We have the direct sum decomposition of the one-graviton subspace
into the direct sum of
$Ran(P)$
and
$Ran(1-P)$.
Let us consider a basis in the one-particle Bosonic subspace formed by a basis
$f_{i}, \quad i \in \N$
of
$Ran(P)$
and a basis
$g_{i}, \quad i \in \N$
of
$Ran(1-P)$.

It is clear that a basis in the $n^{\rm th}$-gravitons subspace is of
the form:
$$
f_{i_{1}} \vee \cdots f_{i_{u}} \vee g_{j_{1}} \vee \cdots \vee g_{j_{v}},
\quad u, v \in \N, \quad u + v = n.
$$

Applying the operator 
$d\Gamma(P)$ 
to such a vector gives the same vector multiplied by
$u$. 
So, in the basis chosen above, the operator 
$d\Gamma(P)$ 
is diagonal with diagonal elements from $\N$. It follows that the operator
$\left. Y\right|_{{\cal H}_{nmls}}$
can also be exhibited into a diagonal form with diagonal elements of the form
$m + l + s + u, \quad u \in \N$.
It is obvious now that for
$m + l + s> 0$
this is an invertible operator.
$\qed$

Now we have a two results which are proved in complete analogy with \cite{Gr1}.

\begin{cor}
Let us define
$
{\cal H}_{0} \equiv \oplus_{n \geq 0} {\cal H}_{n000}
$
and
$
{\cal H}_{1} \equiv \oplus_{n \geq 0, m+l+s > 0} {\cal H}_{nmls}
$.
Then the operator $Y$ has the block-diagonal form
\begin{eqnarray}
Y = \left( \matrix{ Y_{1} & 0 \cr 0 & Y_{0} } \right)
\end{eqnarray}
with
$Y_{1}$
an invertible operator.
\end{cor}

Next, we have the following important proposition.
\begin{prop}
There exists the following vector spaces isomorphism:
\be
Ker(Q)/Im(Q) \simeq {\cal H}'/{\cal H}''
\label{factor}
\ee
where the subspaces
${\cal H}'$
and
${\cal H}''$
have been defined in the previous subsection (see the lemmas \ref{h'1} and
\ref{h''2} respectively).
\label{cohomology}
\end{prop}

{\bf Proof:}
The first two steps of the proof are taken directly from \cite{Gr1} without
changes.

(i) We note that the operators $Q$ and $\tilde{Q}$ have the block-diagonal form
\begin{eqnarray}
Q = \left( \matrix{ Q_{11} & Q_{10} \cr Q_{01} & 0 } \right),
\quad
\tilde{Q} = \left( \matrix{
\tilde{Q}_{11} & \tilde{Q}_{10} \cr \tilde{Q}_{01} & 0 } \right)
\end{eqnarray}
and from the relations (\ref{YQ}) it easily follows that we have
\be
[Y_{1}, Q_{11}] = 0
\label{Y1Q11}
\ee
\be
Y_{1} Q_{10} = Q_{10} Y_{0}, \quad Y_{0} Q_{01} = Q_{01} Y_{1}
\ee
and similar relations for the block-diagonal elements of the homotopy operator
$\tilde{Q}$. In particular we have
\be
[Y_{1}, Q_{10} \tilde{Q}_{01} ] = 0.
\label{Y1Q10}
\ee

(ii) Let now
$\Phi \in Ker(Q)$.
If we apply the relation to (\ref{Y}) the vector $\Phi$ we obtain:
\be
Y \Phi = Q \Psi
\label{Phi}
\ee
where we have defined
\be
\Psi \equiv \tilde{Q} \Phi.
\ee

If we use the block-decomposition form for the vectors $\Phi$ and $\Psi$ we
have in particular, from this relation that $$
\Psi_{0} = \tilde{Q}_{01} \Phi_{1}.
$$

If we use this relation in (\ref{Phi}) we obtain in particular that
\be
Y_{1} \Phi_{1} = Q_{11} \Psi_{1} + Q_{10} \tilde{Q}_{01} \Phi_{1}.
\ee

Because the operator
$Y_{1}$
is invertible, we have from here
\be
\Phi_{1} = Y_{1}^{-1} Q_{11} \Psi_{1} +
Y_{1}^{-1} Q_{10} \tilde{Q}_{01} \Phi_{1}.
\ee

But from (\ref{Y1Q11}) and (\ref{Y1Q10}) we immediately obtain
$$
[Y_{1}^{-1}, Q_{11}] = 0, \quad
[Y_{1}^{-1}, Q_{10}\tilde{Q}_{01}] = 0
$$
so the preceding relations becomes:
\be
\Phi_{1} =  Q_{11} Y_{1}^{-1} \Psi_{1} +
Q_{10} \tilde{Q}_{01} Y_{1}^{-1} \Phi_{1}.
\ee

Now we define the vector $\psi$ by its components:
\be
\psi_{1} \equiv Y_{1}^{-1} \Psi_{1}, \quad
\psi_{0} \equiv \tilde{Q}_{01} Y_{1}^{-1} \Phi_{1}
\ee
and we get by a simple computation
\be
\Phi - Q\psi = \left( \matrix{ 0 \cr \phi_{0} } \right)
\ee
where
$$
\phi_{0} \equiv \Phi_{0} - Q_{01} \psi_{1}.
$$

In other words, if
$\Phi \in Ker(Q)$
then we have
\be
\Phi = Q\psi + \tilde{\Phi}
\ee
where
\be
\tilde{\Phi}^{(nmls)} = 0, \quad m + l +s > 0.
\ee

(iii) The condition
$Q\Phi = 0$
amounts now to
$Q\tilde{\Phi} = 0$
or, with the explicit expression of the supercharge (\ref{Q-explicit}):
\be
q^{\alpha} \tilde{\Phi}^{(n+1,0,0,0)}_{\alpha\sigma,\mu_{1}\nu_{1},\dots,
\mu_{n}\nu_{n};\emptyset;\emptyset}(q,k_{1},\dots,k_{n};\emptyset;\emptyset;
\emptyset) = 0,
\quad \forall n \in \N
\label{transv}
\ee
i.e. the ensemble
$\left. \{\tilde{\Phi}^{(n000)}\}\right|_{n \in \N}$
is an element from
${\cal H}'$
(see lemma \ref{h'1}).

We determine now in what conditions 
$\tilde{\Phi}$
is an element from
$Im(Q)$
i.e. we have
$\tilde{\Phi} = Q \chi$.
It is clear that only the components
$\chi^{(n100)}$
should be taken non-null. Then the expression of the supercharge
(\ref{Q-explicit}) gives for any
$n \in \N$
the following expressions:
\bea
\left(Q \chi\right)^{(n000)}_{\mu_{1}\nu_{1},\dots,\mu_{n}\nu_{n};\emptyset;
\emptyset}(k_{1},\dots,k_{n};\emptyset;\emptyset;\emptyset) =
{1\over 2\sqrt{n}} \sum_{i=1}^{n} 
[(k_{i})_{\mu_{i}} \delta_{\nu_{i}}^{\alpha}
+ (k_{i})_{\nu_{i}} \delta_{\mu_{i}}^{\alpha}
- {1\over 2} k_{i}^{\alpha} g_{\mu_{i}\nu_{i}}] \times
\nonumber \\
\chi^{(n-1,1,0,0)}_{\mu_{1}\nu_{1},\dots,\hat{\mu_{i}}\hat{\nu_{i}},\dots,
\mu_{n}\nu_{n};\alpha;\emptyset;\emptyset}
(k_{1},\dots,\hat{k_{i}},\dots,k_{n};k_{i};\emptyset;\emptyset).
\label{Q-chi}
\eea
The condition
$\tilde{\Phi} = Q \chi$
implies
$\left(Q \chi\right)^{(n110)} = 0$
that's it
\be
q^{\alpha} \chi^{(n+1,1,0,0)}_{\alpha\sigma,\mu_{1}\nu_{1},\dots,
\mu_{n}\nu_{n};\rho;\emptyset}(q,k_{1},\dots,k_{n};p;\emptyset;\emptyset) = 0
\label{1}
\ee
and
$\left(Q \chi\right)^{(n001)} = 0$
which is
\be
r^{\alpha} \chi^{(n,1,0,0)}_{\mu_{1}\nu_{1},\dots,\mu_{n}\nu_{n};\alpha;
\emptyset}(k_{1},\dots,k_{n};r;\emptyset;\emptyset) = 0.
\label{2}
\ee
From (\ref{2}) it follows that the last term from the square bracket in
(\ref{Q-chi}) gives zero and we have in fact:
\bea
\left(Q \chi\right)^{(n000)}_{\mu_{1}\nu_{1},\dots,\mu_{n}\nu_{n};\emptyset;
\emptyset}(k_{1},\dots,k_{n};\emptyset;\emptyset;\emptyset) =
{1\over 2\sqrt{n}} \sum_{i=1}^{n} 
[(k_{i})_{\mu_{i}} \delta_{\nu_{i}}^{\alpha}
+ (k_{i})_{\nu_{i}} \delta_{\mu_{i}}^{\alpha}] \times
\nonumber \\
\chi^{(n-1,1,0,0)}_{\mu_{1}\nu_{1},\dots,\hat{\mu_{i}}\hat{\nu_{i}},\dots,
\mu_{n}\nu_{n};\alpha;\emptyset;\emptyset}
(k_{1},\dots,\hat{k_{i}},\dots,k_{n};k_{i};\emptyset;\emptyset).
\label{Q-chi1}
\eea

If we take 
$
\chi^{(n-1,1,0,0)}
$
to be decomposable i.e. of the form
$
\chi^{(n-1,1,0,0)}(K;p;\emptyset;\emptyset) = \Psi(K) f(p)
$
then from (\ref{1}) it follows that 
$\Psi$
is an element of 
${\cal H}'$
and from (\ref{2}) it follows that the function $f$ verifies the transversality
condition (\ref{trans-f}).  This means that the expression (\ref{Q-chi1}) is an
element from 
${\cal H}_{n}''$ 
(see (\ref{h-secund})).  The isomorphism from the statement is now 
$ 
[\Phi] \leftrightarrow [\tilde{\Phi}] 
$ 
where in the left hand side we take classes modulo $Im(Q)$ and in the right
hand side we take classes modulo ${\cal H}''$.  
$\qed$

We now have some standard results (see \cite{Gr1}):
\begin{lemma}
The sesquilinear form
$(\cdot,\cdot)$
induces a strictly positive defined scalar product on the factor space
$\overline{Ker(Q)/Im(Q)}$.
\label{RR}
\end{lemma}

\begin{lemma}
The representation of ${\cal U}$ of the Poincar\'e group factors out at
$Ker(Q)/Im(Q)$.
\end{lemma}

The main result follows:
\begin{thm}
The isomorphism (\ref{factor}) extends to a Hilbert space isomorphism:
$$
\overline{Ker(Q)/Im(Q)} \simeq {\cal F}_{photon}.
$$
\label{graviton+ghosts}
\end{thm}

We end this Section with a comparison with the alternative formulation of the
quantization process as presented in \cite{KO1} and possible obstacles.
We first note that the commutation relations for the gravitational field
(\ref{CCR-grav-pm}) differ from the similar relations appearing in \cite{KO1}.
So we must first settle this point. Is there a flexibility in the formalism
presented up till now such that one can obtain the commutations relations as
presented in \cite{KO1} ? We will show that this is possible. The idea is the
following one. First, we make the following modifications in proposition
\ref{pre}. We replace ${\sf F}$ by
\be
{\sf F} \equiv \{ h_{\rho\sigma} \vert h_{\rho\sigma} = h_{\sigma\rho} \}
\label{F'}
\ee
and $B$ by
\be
B \equiv \{ (p_{\mu}, h_{\rho\sigma}) \vert p \in X_{0}^{+}, \quad
h_{\rho\sigma} \in {\sf F}, \quad {h_{\rho}}^{\sigma} p_{\sigma} = 0,
\quad {h_{\rho}}^{\rho} = 0 \}
\label{B'}
\ee
that's it we move the condition of tracelessness form the definition of 
${\sf F}$
into the definition of $B$. It is easy to see that the assertions of the 
propositions \ref{pre} and \ref{zero} remain true. Also the assertions of the
proposition \ref{Hilbert} and of the fundamental theorem \ref{graviton} stay
true if we modify appropriately the Borel set
$[B]$, 
namely we take
\be
[B] \equiv \{ (p, [h]) \vert p \in X_{0}^{+}, \quad
[h] \in [{\sf F}], \quad {h_{\rho}}^{\sigma} p_{\sigma} = 0, \quad 
{h_{\rho}}^{\rho} = 0, \quad \forall h \in [h] \}.
\ee

In the same way, we note that lemma \ref{h'} stays true if we take
\be
{\sf H}'\equiv \{ \phi \in {\sf H} \vert \quad p^{\mu} \phi_{\mu\nu}(p) = 0,
\quad {\phi_{\mu}}^{\mu}(p) = 0\}.
\ee

At last, in the definition of Hilbert space
${\cal H}$
(see the beginning of the Section \ref{qgf}) one should give up the
tracelessness condition (c) and redefine
${\cal H}_{n}'$ as follows:
\be
{\cal H}_{n}' = \{ \Phi^{(n)} \in {\cal H}_{n} | \quad k_{1}^{\mu_{1}} 
\Phi^{(n)}_{\mu_{1},\nu_{1};\dots;\mu_{n},\nu_{n}}(k_{1},\dots,k_{n}) = 0,
\quad g^{\mu_{1}\nu_{1}} 
\Phi^{(n)}_{\mu_{1},\nu_{1};\dots;\mu_{n},\nu_{n}}(k_{1},\dots,k_{n}) = 0 \}
\label{h-prim1}
\ee
and the lemmas \ref{h'1}, \ref{h-secund} and the proposition
\ref{graviton-factor} remain true. 

Now we can adopt other definition for the annihilation and creation operators
(\ref{annihilation-grav}) and (\ref{creation-grav}), namely:
\bea
\left( h_{\rho\sigma}(p) 
\Phi\right)^{(n)}_{\mu_{1},\nu_{1};\dots;\mu_{n},\nu_{n}}
(k_{1},\dots,k_{n}) \equiv \sqrt{n+1} 
\nonumber \\
\left[ \Phi^{(n+1)}_{\rho,\sigma;\mu_{1},\nu_{1};\dots;\mu_{n},\nu_{n}}
(p,k_{1},\dots,k_{n}) 
- {1\over 2} g_{\rho\sigma} g^{\lambda\omega}
\Phi^{(n+1)}_{\lambda,\omega;\mu_{1},\nu_{1};\dots;\mu_{n},\nu_{n}}
(p,k_{1},\dots,k_{n})\right],
\quad \forall n \in \N
\label{annihilation-grav1}
\eea
and
\bea
\left( h^{\dagger}_{\rho\sigma}(p) 
\Phi\right)^{(n)}_{\mu_{1},\nu_{1};\dots;\mu_{n},\nu_{n}}
(k_{1},\dots,k_{n}) \equiv {\omega({\bf p}) \over \sqrt{n}}
\sum_{i=1}^{n} \delta({\bf p}-{\bf k_{i}}) 
\left( g_{\rho\mu_{i}} g_{\sigma\nu_{i}} + g_{\rho\nu_{i}} g_{\sigma\mu_{i}} 
- g_{\rho\sigma} g_{\mu_{i}\nu_{i}}\right)
\nonumber \\
\Phi^{(n-1)}_{\mu_{1},\nu_{1};\dots;\hat{\mu_{i}},\hat{\nu_{i}};\dots;
\mu_{n},\nu_{n}} 
(k_{1},\dots,\hat{k_{i}},\dots,k_{n}), 
\qquad \forall n \in \N. \quad
\label{creation-grav1}
\eea

In this case, one can see that everything said in Section \ref{qgf} stays true
with the exception of some of the commutation relation, namely the last
relation (\ref{CCR-grav}), the first relation (\ref{CCR-grav-pm}) and the
relation (\ref{commutation-grav})  which become respectively:
\be
\left[ h_{\rho\sigma}(p), h^{\dagger}_{\lambda\omega}(p')\right] =
\omega({\bf p}) \left( g_{\rho\lambda} g_{\sigma\omega} + 
g_{\rho\omega} g_{\sigma\lambda} - 
{1\over 2} g_{\rho\sigma} g_{\lambda\omega}\right)
\delta{({\bf p}-{\bf p}')} {\bf 1},
\label{CCR-grav1}
\ee
\be
\left[h^{(\mp)}_{\rho\sigma}(x),h^{(\pm)}_{(\lambda\omega)}(y)\right] =
{1\over 2} \left( g_{\rho\lambda} g_{\sigma\omega} + 
g_{\rho\omega} g_{\sigma\lambda} -  g_{\rho\sigma} g_{\lambda\omega}\right)
D^{(\pm)}_{0}(x-y) \times {\bf 1},
\label{CCR-grav-pm1}
\ee
and 
\be
\left[h_{\rho\sigma}(x),h_{\lambda\omega}(y)\right] = 
{1\over 2} \left( g_{\rho\lambda} g_{\sigma\omega} + 
g_{\rho\omega} g_{\sigma\lambda} - g_{\rho\sigma} g_{\lambda\omega}\right)
D_{0}(x-y) \times {\bf 1}.
\label{commutation-grav1}
\ee

So, we have obtained the commutation relations from \cite{KO1}. Now we consider
instead of the auxiliary Hilbert space
${\cal H}^{gh}$
defined by (\ref{Hilbert-ghost}) the subspace corresponding to 
$s = 0$
i.e. we eliminate completely from the game the scalar ghost. The expression of
the supercharge (\ref{supercharge}) becomes
\be
Q \equiv \int_{X^{+}_{0}} d\alpha^{+}_{0}(k) k^{\mu}
\left[ h_{\mu\nu}(k) c^{*\nu}(k) + h^{\dagger}_{\mu\nu}(k) b^{\nu}(k) \right]
\ee
where, of course the operators
$h^{\#}_{\mu\nu}$
have the modified expressions from above; this is exactly the expression from
\cite{KO1}. In this case one can prove that the most important property of the
supercharge, namely (\ref{square}) remains valid. However, we have been unable
to find a corresponding ``homotopy" operator for this supercharge. One can
easily see that an anszatz of the type (\ref{tildeQ}) and (\ref{anszatz-f})
will not gives an operator
$\tilde{Q}$
with desired properties. Even if such an operator could be found, it is
doubtful if we will have a result of the type described in proposition
\ref{cohomology}; indeed, the relation (\ref{transv}) gets modified in our case
into:
\be
q^{\alpha} \tilde{\Phi}^{(n+1,0,0)}_{\alpha\sigma,\mu_{1}\nu_{1},\dots,
\mu_{n}\nu_{n};\emptyset;\emptyset}(q,k_{1},\dots,k_{n};\emptyset;\emptyset) = 
{1\over 2} q_{\sigma} g^{\lambda\omega} 
\tilde{\Phi}^{(n+1,0,0)}_{\lambda\omega,\mu_{1}\nu_{1},\dots,
\mu_{n}\nu_{n};\emptyset}(q,k_{1},\dots,k_{n};\emptyset;\emptyset),
\quad \forall n \in \N
\label{transv1}
\ee
and it is impossible to obtain from this relation the transversality and
tracelessness conditions from the definition of
${\cal H}_{n}'$
appearing in (\ref{h-prim1}). So, it is not clear if the quantum system
described in \cite{KO1} is indeed a system of massless particles of helicity
$2$. 

\newpage
\section{Gauge-Invariant Observables\label{gau-obs}}

We analyse here the construction of observables on the factor space from the
theorem \ref{graviton+ghosts}.

By direct calculus we have the following relations:
\be
\{Q,u_{\mu}^{(\pm)}(x)\} = 0,\quad
\{Q,\tilde{u}_{\mu}^{(\pm)}(x)\} =  - i 
\left[\partial^{\nu} h^{(\pm)}_{\mu\nu}(x) + {1\over 2} \partial_{\mu} 
\Phi^{(\pm)}(x) \right]
\ee
and
\be
[Q, h^{(\pm)}_{\mu\nu}(x)] = {1\over 2}  
\left(\delta_{\mu}^{\rho}\delta_{\nu}^{\sigma} + 
\delta_{\mu}^{\sigma} \delta_{\nu}^{\rho} 
- {1\over 2} g_{\mu\nu} g^{\rho\sigma} \right) 
\partial_{\rho} u_{\sigma}^{(\pm)}(x), \quad
[Q, \Phi^{(\pm)}(x)] = - i \partial^{\rho} u_{\rho}^{(\pm)}(x);
\ee
as a consequence:
\be
\{Q,u_{\mu}(x) \} = 0,\quad
\{Q,\tilde{u}_{\mu}(x) \} =  - i 
\left[\partial^{\nu} h_{\mu\nu}(x) + {1\over 2} \partial_{\mu} 
\Phi(x) \right]
\ee
and
\be
[Q, h_{\mu\nu}(x)] = {1\over 2}  
\left(\delta_{\mu}^{\rho}\delta_{\nu}^{\sigma} + 
\delta_{\mu}^{\sigma} \delta_{\nu}^{\rho} 
- {1\over 2} g_{\mu\nu} g^{\rho\sigma} \right) 
\partial_{\rho} u_{\sigma}(x), \quad
[Q, \Phi(x)] = - i \partial^{\rho} u_{\rho}(x).
\label{Q-com}
\ee

Next, we denote by ${\cal W}$ the linear space of all Wick monomials on the
Fock space
${\cal H}^{gh}$
i.e. containing the fields
$h_{\mu\nu}(x)$,
$u_{\mu}(x)$,
$\tilde{u}_{\mu}(x)$
and
$\Phi(x)$.
If $M$ is such a Wick monomial, we define by
$gh_{\pm}(M)$
the degree in 
$\tilde{u_{\mu}}$ 
(resp. in 
$u_{\mu}$). 
The total degree of $M$ is 
\be
deg(M) \equiv gh_{+}(M) + gh_{-}(M).
\ee

The {\it ghost number} is, by definition, the expression:
\be
gh(M) \equiv gh_{+}(M) - gh_{-}(M).
\ee

If
$M \in {\cal W}$
let us define the operator:
\be
d_{Q} M \equiv :QM: - (-1)^{gh(M)} :MQ:
\label{BRST-op}
\ee
on monomials $M$ and extend it by linearity to the whole ${\cal W}$. Then
$d_{Q}M \in {\cal W}$
and
\be
gh(d_{Q}M) = gh(M) -1.
\ee

The operator
$d_{Q}: {\cal W} \rightarrow {\cal W}$
is called the {\it BRST operator}; other properties of this
object are summarized in the following elementary:
\begin{prop}
The following relations are verified:
\be
d_{Q}^{2} = 0,
\label{Q2}
\ee
\bea
d_{Q} u_{\mu} = 0, \quad 
d_{Q} \tilde{u}_{\mu} = - 
i(\partial^{\nu} h_{\mu\nu} + {1\over 2} \partial_{\mu} \Phi), \quad
d_{Q} \Phi = - i \partial_{\rho} u^{\rho},
\nonumber \\
d_{Q} h_{\mu\nu} = {1\over 2}  
\left(\delta_{\mu}^{\rho}\delta_{\nu}^{\sigma} + 
\delta_{\mu}^{\sigma} \delta_{\nu}^{\rho} 
- {1\over 2} g_{\mu\nu} g^{\rho\sigma} \right) 
\partial_{\rho} u_{\sigma}.
\label{BRST}
\eea
\be
d_{Q}(MN) = (d_{Q}M) N + (-1)^{gh(M)} M (d_{Q}N), \quad
\forall M, N \in {\cal W}.
\label{Leibnitz}
\ee
\end{prop}

Now we have a series of results which are closely analogous to those derived in
\cite{Gr1}.  First, we distinguish a class of observables on the factor space
from theorem \ref{graviton+ghosts}; we have the following result:
\begin{lemma}
If
$O: {\cal H}^{gh} \rightarrow {\cal H}^{gh}$
verifies the condition
\be
d_{Q} O = 0
\label{dQ}
\ee
then it induces a well defined operator
$[O]$
on the factor space
$\overline{Ker(Q)/Im(Q)} \simeq {\cal F}_{photon}$.

Moreover, in this case the following formula is true for the matrix elements of
the factorized operator
$[O]$:
\be
([\Psi], [O] [\Phi]) = (\Psi, O \Phi).
\label{matrix-elem}
\ee
\end{lemma}

This kind of observables on the physical space will also be called {\it gauge
invariant observables}. Next, we have:
\begin{lemma}
An operator
$O: {\cal H}^{gh} \rightarrow {\cal H}^{gh}$
induces a gauge invariant observables if and only if it verifies:
\be
\left. d_{Q} O \right|_{Ker(Q)} = 0.
\ee
\label{gi-ghosts}
\end{lemma}

Not all operators verifying the condition (\ref{dQ}) are interesting. In fact,
we have from (\ref{Q2}):
\begin{lemma}
The operators of the type
$d_{Q} O$
are inducing a null operator on the factor space; explicitly, we have:
\be
[d_{Q} O] = 0.
\ee
\label{trivial}
\end{lemma}

Moreover, we have:
\begin{thm}
Let the interaction Lagrangian be a Wick monomial
$T_{1} \in {\cal W}$
with
$gh(T_{1}) \not= 0$.
Then the chronological product are null, i.e. there is no non-trivial
$S$-matrix.
\label{gh(M)=0}
\end{thm}

In the framework of pertubative quantum field theory the axiom of factorization
in the adiabatic limit is:
\be
\lim_{\epsilon \searrow 0} d_{Q} \int_{\R^{4}} dx 
\left. T_{n}(x_{1},\dots,x_{n}) \right|_{Ker(Q)} = 0,
\quad \forall n \in \N^{*}.
\label{factorization1}
\ee

If infrared divergences cannot be avoided, the one can consider the preceding
relation at the heuristic level and impose the postulate:
\be
d_{Q} T_{n}(x_{1},\dots,x_{n}) = i \sum_{l=1}^{n}
{\partial \over \partial x^{\mu}_{l}} 
T^{\mu}_{n/l}(x_{1},\dots,x_{n}),
\quad \forall n \in \N^{*}
\ee
as it is done in \cite{Sc1}, \cite{DHKS1}.

\end{document}